\documentclass[a4paper,11pt]{article}
\usepackage{mathtools,amssymb}
\usepackage{bm}
\usepackage{caption}
\usepackage{subcaption}
\usepackage[round]{natbib}
\usepackage{rotating}

\renewcommand{\thefootnote}{\fnsymbol{footnote}}

\title{Natural Groundwater Systems Can Display Chaotic Mixing at the Darcy Scale}

\author{M. G. Trefry$^{1}$\thanks{Corresponding author M.G. Trefry: michaelgtrefry@gmail.com}, 
        D. R. Lester$^2$, Guy Metcalfe$^{3}$, J. Wu$^{2}$\\
  $^1$Independent Researcher\\
  $^2$School of Engineering, RMIT\\
  $^3$School of Engineering, Swinburne University of Technology\\
  Australia}

\date{2018-Aug-08}

\begin{document}

\maketitle

\noindent{\bf Key Points:}
\begin{itemize}
\item Time-periodic Darcy flows in heterogeneous compressible aquifers
  can generate chaotic mixing dynamics
\item Widespread occurrence of coherent Lagrangian structures is
  controlled by key parameter groups and results from common physical
  processes
\item Such structures fundamentally change our view of flow,
  transport, mixing and reaction in groundwater discharge systems
\end{itemize}

\begin{abstract}
  Although steady, isotropic Darcy flows are inherently laminar and
  non-mixing, it is well understood that transient forcing via
  engineered pumping schemes can induce rapid, chaotic mixing in
  groundwater.  In this study we explore the propensity for such
  mixing to arise in natural groundwater systems subject to cyclical
  forcings, e.g. tidal or seasonal influences.  Using a conventional
  linear groundwater flow model subject to tidal forcing, we show that
  under certain conditions these flows generate Lagrangian transport
  and mixing phenomena (chaotic advection) near the tidal boundary.
  We show that aquifer heterogeneity, storativity, and forcing
  magnitude cause reversals in flow direction over the forcing cycle
  which, in turn, generate coherent Lagrangian structures and chaos.
  These features significantly augment fluid mixing and transport,
  leading to anomalous residence time distributions, flow segregation,
  and the potential for profoundly altered reaction kinetics.  We
  define the dimensionless parameter groups which govern this
  phenomenon and explore these groups in connection with a set of
  well-characterised coastal systems.  The potential for Lagrangian
  chaos to be present near discharge boundaries must be recognized and
  assessed in field studies.
\end{abstract}

\newpage

\renewcommand{\thefootnote}{\arabic{footnote}}

\section{Introduction}
\label{sec:intro}

For well over a century it has been known that steady isotropic Darcy
flows are inherently poor mixing flows.  Seminal works by
\cite{Kelvin:1884aa} and \cite{Arnold:1965} established that the
streamlines of these flows are confined to two-dimensional (2D)
lamellar sheets (thus termed ``complex lamellar'' flows), which
significantly restricts their transport behaviour.  This kinematic
constraint is a direct consequence of the helicity-free nature of
Darcy flow, in that the helicity density of these flows (defined as
the dot product of vorticity and velocity~\citep{Moffatt:1969aa}),
which characterises helical twisting of streamlines, is identically
zero, confining streamlines to coherent lamellar sheets.  Confinement
restricts stretching of fluid elements to be at most algebraic in
time~\citep{Dentz:2016aa}, limiting growth of material interfaces and
hence mixing.  Steady isotropic Darcy flows may be strongly
heterogeneous, but they exhibit slow (algebraic) deformation of fluid
elements and poor mixing.

Conversely, transient isotropic Darcy flows break this kinematic
constraint as the confining lamellae can change with time, leading to
the possibility of \emph{chaotic mixing} which is characterised by
exponential stretching of fluid elements~\citep{Ottino:1989aa}.  Over
the last decade it has been established that such rapid mixing may be
\emph{engineered} at the Darcy scale (metres to kilometres) in
groundwater systems through the use of programmed pumping activities
\citep[][]{Sposito:2006aa,Bagtzoglou:2007aa,Metcalfe:2007ab,Lester:2009ab,Lester:2010aa,Metcalfe:2010aa,Metcalfe:2010ab,Metcalfe:2011aa,Trefry:2012aa,Mays:2012,Piscopo:2013aa,Rodriguez:2017aa,Cho:2017aa},
with selectable and comprehensive consequences for contaminant
migration and fate.  The saturated zone may now be regarded as a
reaction vessel whose mixing state may be \emph{manipulated} in
order to promote reactivity or segregation according to the
requirements of groundwater quality management.  This is a significant
paradigm shift for groundwater systems which have long been thought to
be governed by poor mixing and transport characteristics unamenable to
external control.

Now that we can engineer chaotic mixing in groundwater systems, we
may consider the supplementary question, is chaotic mixing a
\emph{natural characteristic} of groundwater systems? If so, the
ramifications for flow, transport and reaction in groundwater systems
are likely to be as profound as those demonstrated elsewhere for
chaotic flows in diverse areas of science (see e.g. \citet{Tel2005} or
\citet{Aref_frontiers_2017}).  In order to answer this question we
seek natural (unmodified by human intervention) and common aquifer
settings that engender the basic preconditions for Lagrangian chaos,
namely repeated stretching and folding of fluid elements, widely
considered a hallmark of chaotic advection~\citep{Ottino:1989aa}.  If
we can demonstrate that such conditions occur in natural settings, and
we can quantify how these lead to chaotic mixing, then we will have
established that such kinematics are a natural phenomenon in a broad
class of groundwater systems.  This demonstration is the purpose of
this paper.

This question has been partially answered in the context of steady \emph{anisotropic} Darcy flows, which do not conform to the kinematic constraints of isotropic Darcy flow (they are not helicity-free) and have been shown both numerically~\citep{Cirpka:2012aa} and experimentally~\citep{Ye:2015aa} to exhibit the accelerated mixing dynamics characteristic of chaotic mixing. In this study we determine to what extent \emph{transient forcings} can generate chaotic mixing in natural groundwater systems. Many natural groundwater systems are driven by discharge boundaries at rivers, lakes and coasts, which are always time-dependent to some degree through tidal action, seasonal, and barometric effects etc. In principle,  transient forcings are sufficient to break the zero helicity kinematic constraint, but to generate chaotic mixing at the Darcy scale these must also lead to stretching and folding fluid motions: changes in flow orientation and reorganisation of the flow streamlines are necessary to achieve persistent stretching and folding  of material elements. As such, natural groundwater systems near discharge boundaries represent likely candidates for chaotic mixing because the interaction between transient forcing at the discharge boundary combined with spatial heterogeneity of aquifer properties (intrinsic to all groundwater systems) provides a mechanism for strong flow reorientation over the forcing cycle. Indeed, these physical phenomena influence the transport and fate of groundwater contaminants at the intertidal zones, leading to complex spatio-temporal distributions of biogeochemical reactivity \citep[][]{Heiss:2017aa,Liu:2017aa,Malott:2017aa,Kobayashi:2017aa}. Fluid mixing, dilution and residence time have been correlated with enhanced biodegradation rates for contaminants in coastal aquifers subject to tidal influences \citep[][]{Robinson:2009aa,Geng:2017aa}. Since chaotic mixing can fundamentally alter reaction kinetics, leading to e.g. singularly enhanced reaction rates \citep{Toroczkai:1998aa}, we hypothesize that chaotic processes may contribute to groundwater discharge and reaction complexity in coastal zones. Accordingly, for the remainder of this paper we focus on coastal aquifers as likely candidates for natural chaotic mixing.

In pursuing this approach we seek to reduce the complexity of the
physical model in order to emphasize the source and nature of key
chaotic processes.  Thus, we are concerned with confined, isohaline
systems (zero density differences) in two dimensions (plan view),
resulting in a linear flow equation within a simple rectangular
domain.  Issues of capillarity and beach slope are ignored in this
study, as are solute transport and dispersion.  These assumptions are
contrary to established theoretical approaches for studying
groundwater discharge and reaction in coastal systems; however, our
contention is that if conditions are right for chaotic signatures to
occur in our simple model tidal flow system, then such signatures are
also possible in more realistic models conditioned on field
observations.

We proceed as follows. After a brief survey of chaotic fluid processes in porous media, the paper introduces a conventional model 2D Darcian system representing a heterogeneous, tidally forced confined aquifer that is used as a basis for a Lagrangian analysis of the flow regime. The tidal flow problem is solved numerically and the basic solution properties are explored, focusing on the flux and velocity distributions. In subsequent sections key dimensionless parameter groups are introduced to assist in predicting the onset of chaotic mixing, a survey of Lagrangian and chaotic phenomena arising in the model tidal system is presented, and comments are made on the potential ramifications for biogeochemical processes acting at tidal discharge boundaries.

\section{Chaotic advection in porous media flows}
At the pore scale, fluid advection and mixing within the pore space $\Omega$ is governed by the Stokes equation (\ref{eq:stokes}) subject to no-slip conditions at the fluid boundary $\partial\Omega$
\begin{align} \label{eq:stokes}
\mu  \nabla^{2} \mathbf{v}(\mathbf{x}) - \nabla p(\mathbf{x}) = \mathbf{0},\quad\nabla \cdot \mathbf{v}(\mathbf{x}) = 0,\quad\mathbf{x}\in\Omega, \quad \mathbf{v}(\mathbf{x})\big|_{\mathbf{x}\in\partial\Omega}=0,
\end{align}
where $\mu$ is the dynamic viscosity, $p$ is the fluid pressure,
$\nabla^{2}$ is the vector Laplacian operator, and $\mathbf{v}$ is the
fluid velocity vector of the pure fluid within the pore space
$\Omega$.  This boundary-dominated flow is controlled by the geometry
and connectivity of the pore space.  Three-dimensional steady Stokes
flows engender chaotic mixing due to the topological complexity
inherent to all porous media~\citep{Lester:2013aa, Lester:2015aa}, and
the resulting pore-scale transport signatures persist into the
macroscale~\citep{Lester:2016ad, Lester:2014aa}.

Whilst these models provide insights into dispersion and mixing at the
pore scale, continuum flow models that utilise averaged microscale
pore and fluid properties are more widely used.  These models are
employed in the analysis of macroscale systems for which observational
data are often scarce and the difficulties of microscale system
characterisation and upscaling are usually insurmountable.  In the
groundwater domain, the Darcy flow equation
\citep[][]{Darcy:1856aa,Whitaker:1986aa} is a useful continuum model,
relating fluid flux through a porous medium to imposed pressure
gradient.  Despite the ubiquity of pore-scale chaotic mixing, Darcy
flows are laminar and poorly mixing due to the zero helicity kinematic
constraint.  It was not recognised until recently
\citep[][]{Sposito:2006aa,Bagtzoglou:2007aa,Metcalfe:2007ab, Lester:2009ab,
  Metcalfe:2010ab,Trefry:2012aa,Piscopo:2013aa} that Darcian systems
could display chaotic signatures which underpin mixing enhancement
which may be beneficial for contaminant remediation
\citep{Rodriguez:2017aa}.  The mechanism generating macroscale chaotic
mixing is the ``crossing" of streamlines (when viewed at different
times) which can lead to stretching and folding of the flow. Repeated
changes of fluid flow direction can enhance reaction in porous media \citep{Zhang:2009aa} and, critically, produce kinematic effects
\citep{Trefry:2012aa} ranging from the creation of kinematic transport
``barriers'' to rapid, global mixing.  A key hydrogeological example
is the programmed \emph{rotated potential mixing} (RPM) flow
\citep{Lester:2009ab,Metcalfe:2010aa} which can be generated in the
field by sets of injection/extraction wells operating under a
synchronised dipole pumping schedule \citep{Cho:2017aa}.
Three-dimensional variants of the RPM flow have also been
studied~\citep{Smith:2016aa,Cho:2017aa}, showing controllable 3D
chaotic mixing is  possible in porous media flows at the Darcy
scale.

For natural Darcy flows we need a quantitative, computable metric that
definitively indicates whether chaos exists in the flow or not.  The
Lyapunov exponent is commonly used in dynamical systems to measure the
sensitivity of a system to its initial conditions; it directly
characterises the rate of exponential growth of (infinitesimal)
material elements \citep[see for example][]{Allshouse:2015aa} in both
volume-preserving flows and more recently, non-volume-preserving flows
\citep[][]{Volk:2014aa,Gonzalez:2016aa,Perez:2014aa}.  
The time-dependent groundwater equation (\ref{eq:gfe}) describes a
non-volume-preserving flow due to the storage term, and typically
corresponds to an \emph{open flow} \citep{Tel2005} due to the in-
and outflow boundaries typical of hydrogeological models.  In this
study we use the presence of a positive Lyapunov exponent (indicating
the presence of exponential fluid stretching) as an indicator of
chaotic mixing.  We now turn to a formal problem definition for our
tidally forced system.

\section{Transient flow in tidally forced aquifers}

\begin{figure}[t]
\centering
\includegraphics[scale=0.35]{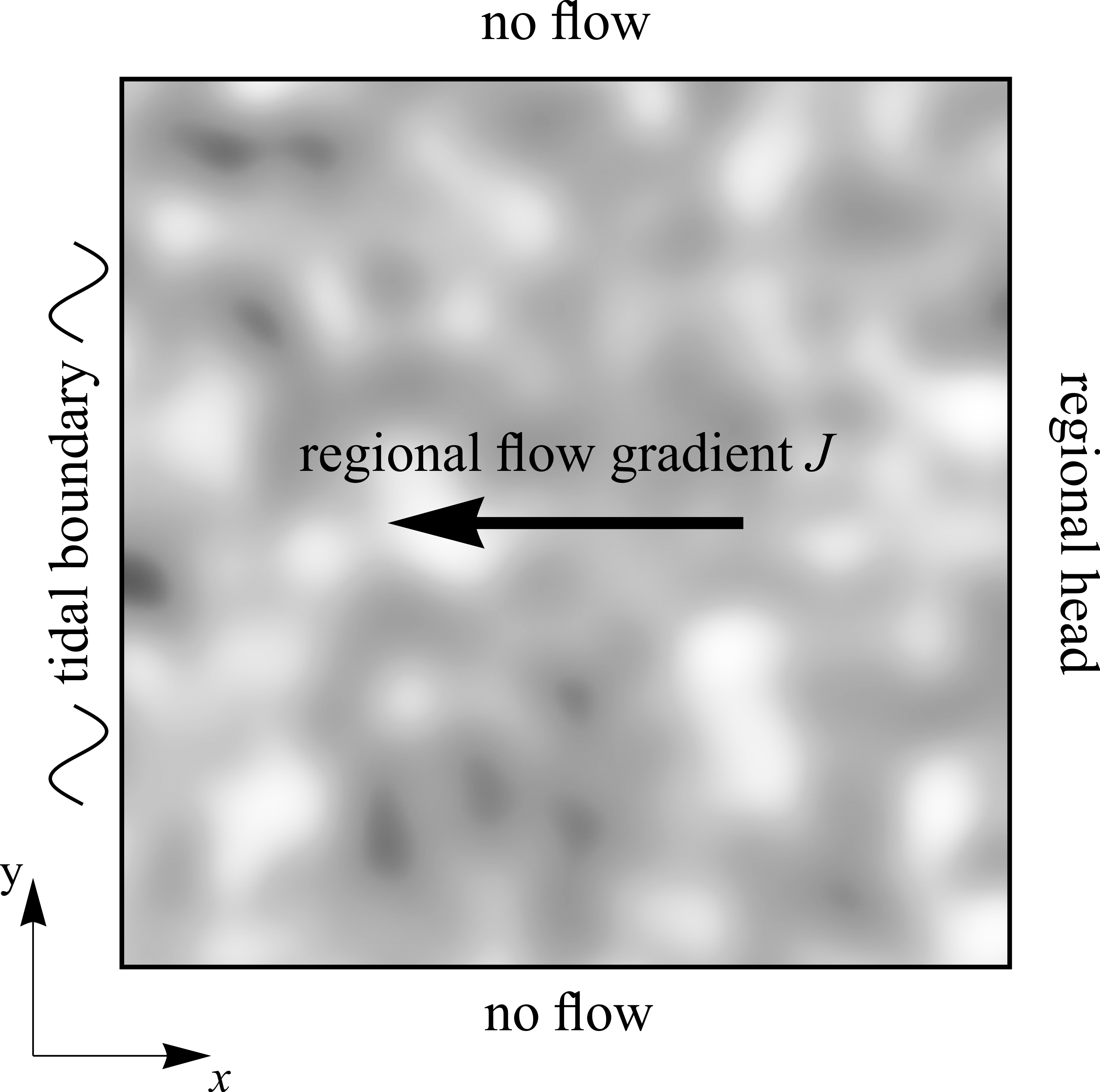}
\caption{Schematic of the square aquifer domain $\mathcal{D}$ and
  boundaries, showing the regional flow gradient $J$ and spatial
  heterogeneity of $K$.}
\label{fig:domain}
\end{figure}

As a prototype of a coastal aquifer, we consider a bounded 2D domain in the $x-y$ plane representing an aquifer in plan view, that is the aquifer domain $\mathcal{D}$ is defined by the coordinate $\textbf{x} = (x,y) \in [0,L_{x}]\times[0,L_{y}]$. We assume the aquifer has unit thickness $B = 1$. For numerical simplicity and without loss of generality we assume $L_{x} = L_{y} = L$, as depicted in Figure \ref{fig:domain}. Furthermore we limit our analysis to confined flow conditions in the absence of vertical recharge and internal sources and sinks. Following \citet{Bear:1972aa} we write the continuity equation for an incompressible fluid in a deformable porous medium of matrix porosity $\varphi$ as
\begin{equation} \label{eq:cedeform}
\dfrac{\partial \varphi}{\partial t}+\nabla\cdot\mathbf{q}= 0,
\end{equation}
where the Darcy flux $\mathbf{q}=\varphi\mathbf{v}$ (with $\mathbf{v}$
the groundwater velocity) is given by the Darcy equation
\begin{equation} \label{eq:Darcyq}
\textbf{q} = -K \nabla h,
\end{equation}
where the spatially heterogeneous scalar $K(\mathbf{x})$ is the isotropic saturated hydraulic conductivity and $h$ is the pressure head. For 2D systems it is usual to work in terms of the transmissivity $T = K B$; however, as $B=1$ we use $K$ throughout. Following conventional approaches~\citep{Bear:1972aa,Coussy:2004aa}, we model changes in the local porosity $\varphi$ due to fluctuations in the local head $h$ via a linear approximation 
\begin{equation} \label{eq:nvsh}
\varphi(h) = \varphi_{\textrm{ref}} + S\!_{s} \; (h - h_{\textrm{ref}}),
\end{equation}
where $h_{\textrm{ref}}$ is the reference head at which the reference porosity $\varphi_{\textrm{ref}}$ applies, and $S\!_{s}$ is formally the specific storage. However, noting the unit aquifer thickness assumption we henceforth replace $S\!_{s}$ by $S$ and refer to the storage term simply as storativity. As $\varphi_{\textrm{ref}}$, $h_{\textrm{ref}}$ and $S$ are assumed constant, all spatial and temporal dependence of $\varphi$ is generated by the coupling with $h$. It is important to note that whilst the solid phase in most aquifers is essentially incompressible, equation (\ref{eq:nvsh}) models changes in local porosity with head as an effective compressibility. This conventional approximation avoids the complication of explicitly solving the solids phase displacement~\citep{Bear:1972aa}, but does not fully capture the migration of solids in the aquifer due to pressure gradients. Combining (\ref{eq:cedeform})-(\ref{eq:nvsh}) yields the linear groundwater flow equation
\begin{equation} \label{eq:gfe}
S \dfrac{\partial h}{\partial t} = \nabla\cdot(K \nabla h),
\end{equation}
subject to the no flow, inland fixed head ($J$ the inland head
gradient) and tidal boundary conditions ($g(t)$), respectively,
\begin{align} \label{eq:boundaryconds}
\dfrac{\partial h}{\partial y}\Bigr|_{y=0}=\dfrac{\partial h}{\partial y}\Bigr|_{y=L} = 0, & & h(L,y,t) = J L, & & h(0,y,t)=\textsl{g}(t). 
\end{align}
Solution of the groundwater equation (\ref{eq:gfe}) subject to the
boundary conditions (\ref{eq:boundaryconds}) completely solves the
system, and the Darcy flux $\mathbf{q}$ is computed via
(\ref{eq:Darcyq}).  The inland boundary condition $J L$ is intended to
provide for mean discharge flow to the tidal boundary, i.e.
$J L > \overline{\textsl{g}(t)}$, which is common in the field.
Nevertheless, saline intrusion $J L < \overline{\textsl{g}(t)}$ is
becoming more widespread \citep[see, e.g.][]{Fadili:2018aa}.  Our
model encompasses mean intrusion from the tidal boundary, but in this
work we restrict attention solely to discharging regional flows.

\subsection{Steady and periodic solutions}

Following \citet{Trefry:2011aa}, we assume the tidal boundary forcing
function $\textsl{g}(t)$ to be finite-valued and cyclic with period
$P$, i.e. $\textsl{g}(t) = \textsl{g}(t+P)$.  For simplicity of
exposition we assume the tidal forcing to consist of a single Fourier
mode
\begin{equation} \label{eq:fourier}
\textsl{g}(t) = \textsl{g}_p e^{i \omega t },
\end{equation}
where $\omega$ is the forcing frequency, and note that extension to a
multi-modal tidal forcing spectrum does not alter qualitative aspects
of the problem~\citep{Trefry:2004aa}.  Under such forcing, the head
$h$ can be decomposed into steady $h_s$ and periodic $h_p$ components
\begin{equation} \label{eq:hdecomp}
h(\textbf{x},t) = h_{s}(\textbf{x}) + h_{p}(\textbf{x},t) = h_{s}(\textbf{x}) + h_{p,x}(\textbf{x})e^{i\omega t}
\end{equation}
that satisfy the steady and periodic Darcy equations, respectively,
\begin{align} \label{eq:Helmholtz}
\nabla\cdot(K \nabla h_{s}(\mathbf{x})) = 0, & & \nabla\cdot(K \nabla h_{p,x}(\mathbf{x})) - i \omega S h_{p,x}(\mathbf{x}) = 0,
\end{align}
subject to the boundary conditions
\begin{align} \label{eq:hmbc}
h_{s}(L,y) = J L, & & h_{s}(0,y) =0, & & h_{p,x}(L,y) = 0, & & h_{p,x}(0,y)& =\textsl{g}_p
\end{align}
with zero flux conditions for both $h_s$, $h_{p,x}$ at the $y=0$ and
$y=L$ boundaries.  All periodic quantities are complex in this
formulation, and so the real part must be taken throughout as these
are observable quantities.  Henceforth we drop the notations $h_{p,x}$
for the spatial part of the periodic head component, replacing it by
$h_{p}$.  It is understood that formally
$h_{p}(\mathbf{x},t) \equiv h_{p}(\mathbf{x}) e^{i\omega t}$.
Likewise, the porosity relation (\ref{eq:nvsh}) can be decomposed into
steady and periodic contributions as
\begin{align} \label{eq:nvshperiodic}
\varphi(\mathbf{x},t) =\varphi_s(\mathbf{x})+\varphi_p(\mathbf{x},t)=\varphi_{\textrm{ref}} + S (h_{s}(\mathbf{x}) - h_{\textrm{ref}}) + S h_{p}(\mathbf{x},t),
\end{align}
where $\varphi_p$ is the last term on the right hand side of
(\ref{eq:nvshperiodic}).

\subsection{Hydraulic characteristics} \label{subsec:hydraulic_chars}

Early analytical results for tidal influences in one-dimensional
aquifers with uniform and homogeneous $K$ fields were derived for
semi-infinite domains by \citet{Jacob:1950aa} and were extended to
finite \citep{Townley:1995aa}, layered \citep{LiJiao:2002aa} and
composite \citep{Trefry:1999aa} domains.  Analytical results were also
obtained in two dimensions for uniform \citep{LiBarry:2000aa} and stochastic
\citep{Trefry:2011aa} aquifer property distributions.  Many other
analytical results for tidal groundwater systems have been reported.

A key feature of the tidal solutions is a finite propagation speed from the tidal boundary to the interior, so that induced oscillations measured at locations inside the aquifer domain are lagged (out of phase) and attenuated (reduced in amplitude) with respect to the boundary forcing condition. For a one-dimensional, semi-infinite homogeneous aquifer, the phase lag $\Delta\tau$ and attenuation $\alpha$ are related to the forcing frequency $\omega$, aquifer diffusivity $D=K/S$ and distance $x$ from the tidal boundary as \citep[][]{Jacob:1950aa,Ferris:1951aa}
\begin{align} \label{eq:Jacob}
 \Delta\tau = x/\sqrt{2\omega D}, & &\alpha = \textrm{exp}\left(-x \sqrt{\omega/2D}\right).
\end{align}
The dissipative nature of the
periodic equation (\ref{eq:Helmholtz}) provides an exponential decay
of oscillation amplitude with distance from the forcing boundary and
phase lag increasing linearly with distance, with the attenuation
greatest for high $\omega$ and least for low $\omega$.  The Electronic
Supplementary Material accompanying this paper contains example
animations of the solution head distribution, velocity ellipses
(Section \ref{subsec:flux_ellipses}) and the vorticity (Appendix
\ref{subsec:vorticity}).  The animations show the finite propagation
speed of head disturbances in the aquifer domain and the effect of $K$
heterogeneity.

\subsection{Flux ellipses and flow reversal} \label{subsec:flux_ellipses}

\begin{figure} 
\centering
\includegraphics[scale=0.3]{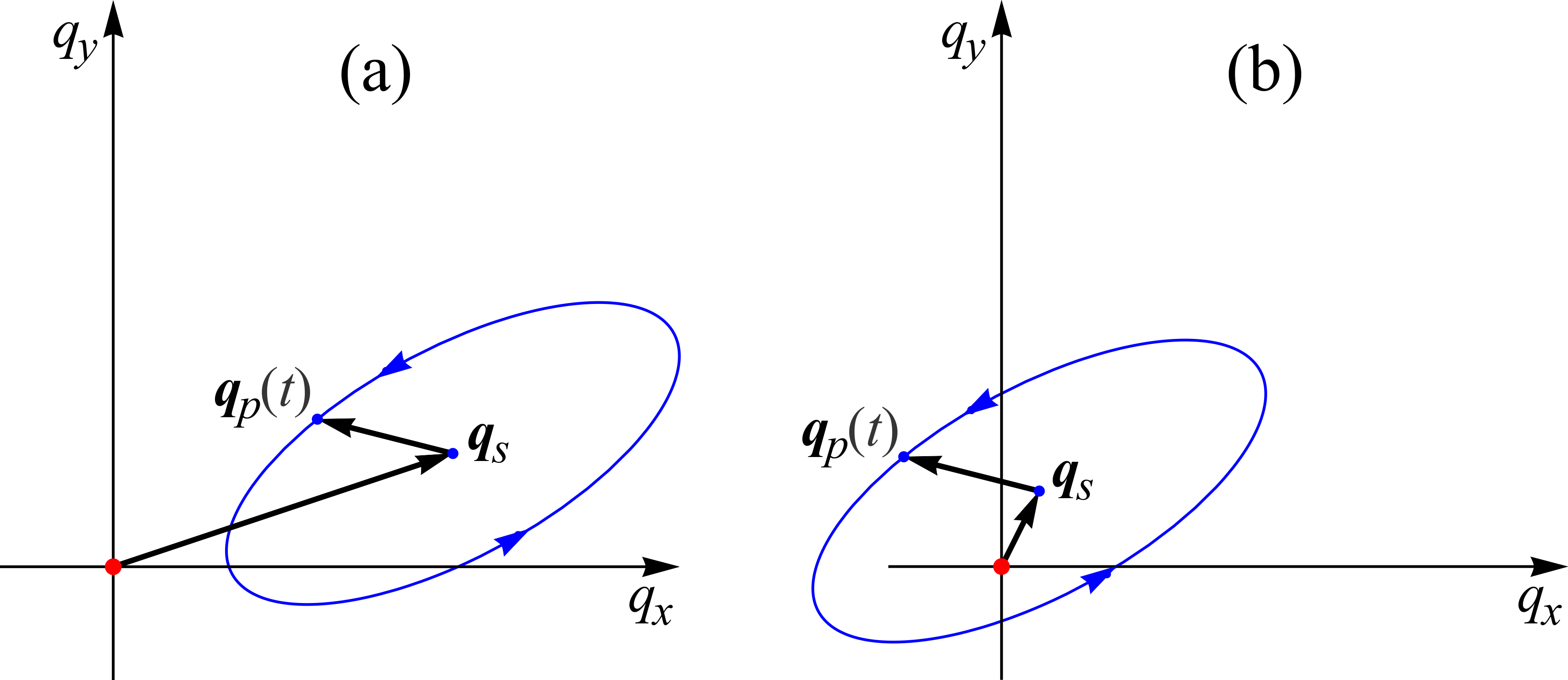}
\caption{Schematic of flux ellipses drawn in blue - (a) regular (excluding the origin), and (b) canonical (containing the origin). Arrows on the ellipses indicate the direction of increasing time $t$. Both flux components of the canonical ellipse change sign at times during the oscillation period.}
\label{fig:qellipse}
\end{figure}

As shown by \citet{Kacimov:1999aa} and \citet{Smith:2005aa}, time-periodic groundwater forcing causes elliptical velocity orbits at each point in the aquifer domain. More precisely, the family of possible velocity orbits includes ellipses, circles, lines and points, depending on degeneracies in the semi-axes and component phases of the underlying flux ellipses. Returning to the present tidal problem, the induced Darcy flux $\mathbf{q}(\mathbf{x},t) = (q_{x}, q_{y})$ at a location $\textbf{x} = (x,y)$ and time $t$ can also be decomposed into steady and periodic components as
\begin{equation} \label{eq:periodicq}
 \mathbf{q}(\mathbf{x},t) = \mathbf{q}_{s}(\mathbf{x})+\mathbf{q}_{p}(\mathbf{x},t) \equiv \mathbf{q}_{s}(\mathbf{x})+\mathbf{q}_{p}(\mathbf{x})e^{i\omega t},
\end{equation}
where $\mathbf{q}_s = -K\nabla h_s$, $\mathbf{q}_{p} = -K\nabla h_{p}$. For any location $\mathbf{x}$ in the flow domain, the steady flux $\mathbf{q}_s(\mathbf{x})$ fixes the location of the flux ellipse shown in in Figure \ref{fig:qellipse}, whereas the periodic flux $\mathbf{q}_{p}(\mathbf{x},t)$ governs the orientation, eccentricity and magnitude of the modal flux ellipse that is swept out parametrically with $t$ over a flow period.   The sign of $\mathbf{q}_{p}$ dictates whether the flux ellipse is traced out clockwise or anti-clockwise in time. For more complex multi-modal tidal forcing $\textsl{g}(t)$, the orbit traced out by $\mathbf{q}$ may be more complex than a simple ellipse but the qualitative aspects of the system are conserved.

Under the condition that the magnitude of the periodic flux is greater than that of the steady flux at a given point $\mathbf{x}$, i.e. $||\mathbf{q}_{p}(\mathbf{x})||>||\mathbf{q}_s(\mathbf{x})||$, then \emph{flow reversal} occurs at some time in the flow cycle, i.e. the total flux $\mathbf{q}$ is in the opposite direction to the steady flux $\mathbf{q}_s$, and the flux ellipse over the entire flow cycle contains the origin $\mathbf{q}=\mathbf{0}$. Flux ellipses that contain the origin (and hence admit flow reversal) will henceforth be referred to as \emph{canonical}. In Appendix~\ref{subsec:vorticity} we show that flow reversal occurs due to the interplay between conductivity variations and compressibility of the aquifer.

As shown in Appendix~\ref{app:zero_compression}, for incompressible
aquifers ($S=0$) the periodic flux vector $\mathbf{q}_p(\mathbf{x},t)$
aligns with the steady flux vector $\mathbf{q}_s(\mathbf{x})$ and the
sign and magnitude of $\mathbf{q}_p$ simply oscillates over a tidal
forcing cycle. Consequently the flux ellipses in the incompressible limit
have zero width (or infinite eccentricity). This leads to fluid
particle trajectories that follow simple, smooth streamlines (although
particles move backwards and forwards as they propagate along
these streamlines), leading to regular, non-chaotic fluid motion. Accordingly, we denote any flux ellipse with an eccentricity greater
than 100 as a \emph{trivial} ellipse as it will not contribute
significantly to complex fluid motion.

\subsection{Fluid velocity and Lagrangian kinematics}

As fluid particles and passive tracers are advected by the groundwater
velocity $\mathbf{v}=\mathbf{q}/\varphi$, in this study we are
primarily interested in the mixing and transport properties of
$\mathbf{v}(\mathbf{x},t)$.  For finite $S>0$, $\varphi$ varies with space and time
through its dependence on $h$, and it is not possible to separate
$\textbf{v}$ into steady and periodic terms; rather, $\mathbf{v}$ is
given by
\begin{equation} \label{eq:vsingle}
 \mathbf{v}(\mathbf{x},t) = \dfrac{\mathbf{q}(\mathbf{x},t)}{\varphi(\mathbf{x},t)} = \dfrac{\mathbf{q}_{s}(\mathbf{x}) + \mathbf{q}_{p}(\mathbf{x},t)}{\varphi_{s}(\mathbf{x}) + \varphi_{p}(\mathbf{x},t)}.
\end{equation}
This velocity formulation is consistent with (\ref{eq:gfe}) where the compression term acts only on net storage; head-dependent alterations to the conductivity $K$ are also plausible but here are neglected in the context of confined aquifers characterized by small $S$. 

Fluid mixing and transport are direct properties of the \emph{Lagrangian kinematics} of the aquifer, which describe the evolution of non-diffusive, passive tracer particles advected by the velocity field as
\begin{equation} \label{eq:pathline}
\dfrac{d\textbf{x}}{dt} = \textbf{v}(\textbf{x},t).
\end{equation}
Whilst the advection equation (\ref{eq:pathline}) appears simple, under certain conditions it
can give rise to distinct regions with chaotic particle trajectories; i.e. solutions to (\ref{eq:pathline}) can exhibit chaotic dynamics. These chaotic regions may be interspersed with regions of regular transport (e.g.\/ smooth,
regular streamlines), defining the \emph{Lagrangian topology} of the flow.  Resolution and classification of these diverse regions may provide deep insights into the mixing and transport properties of aquifer flow.

We use (\ref{eq:pathline}) extensively to resolve the Lagrangian
kinematics and topology of the aquifer flow.  Whilst it appears
incongruous to study mixing in the absence of particle diffusion or
dispersion, we deliberately omit particle diffusion as we wish to
clearly observe the Lagrangian kinematics and Lagrangian topology in
the absence of diffusive noise.  Throughout this study we use the term
``mixing'' to denote the mixing of particle trajectories (akin to the
mixing of coloured balls) rather than the classical definition made in
terms of a decrease in concentration variance or increase in
concentration entropy~\citep{Kitanidis:1994aa}. This is the dynamical
systems perspective of mixing \citep{Aref_frontiers_2017}; Lagrangian
measures give an \emph{advective template} for fluid transport on
which diffusion may subsequently be added~\citep{Lester:2014ab}.

\subsection{Impact of flow reversal on transport and mixing}
\label{subsec:flow_reversal}
\begin{figure}[p]
\begin{centering}
\begin{tabular}{cc}
\includegraphics[width=0.52\columnwidth]{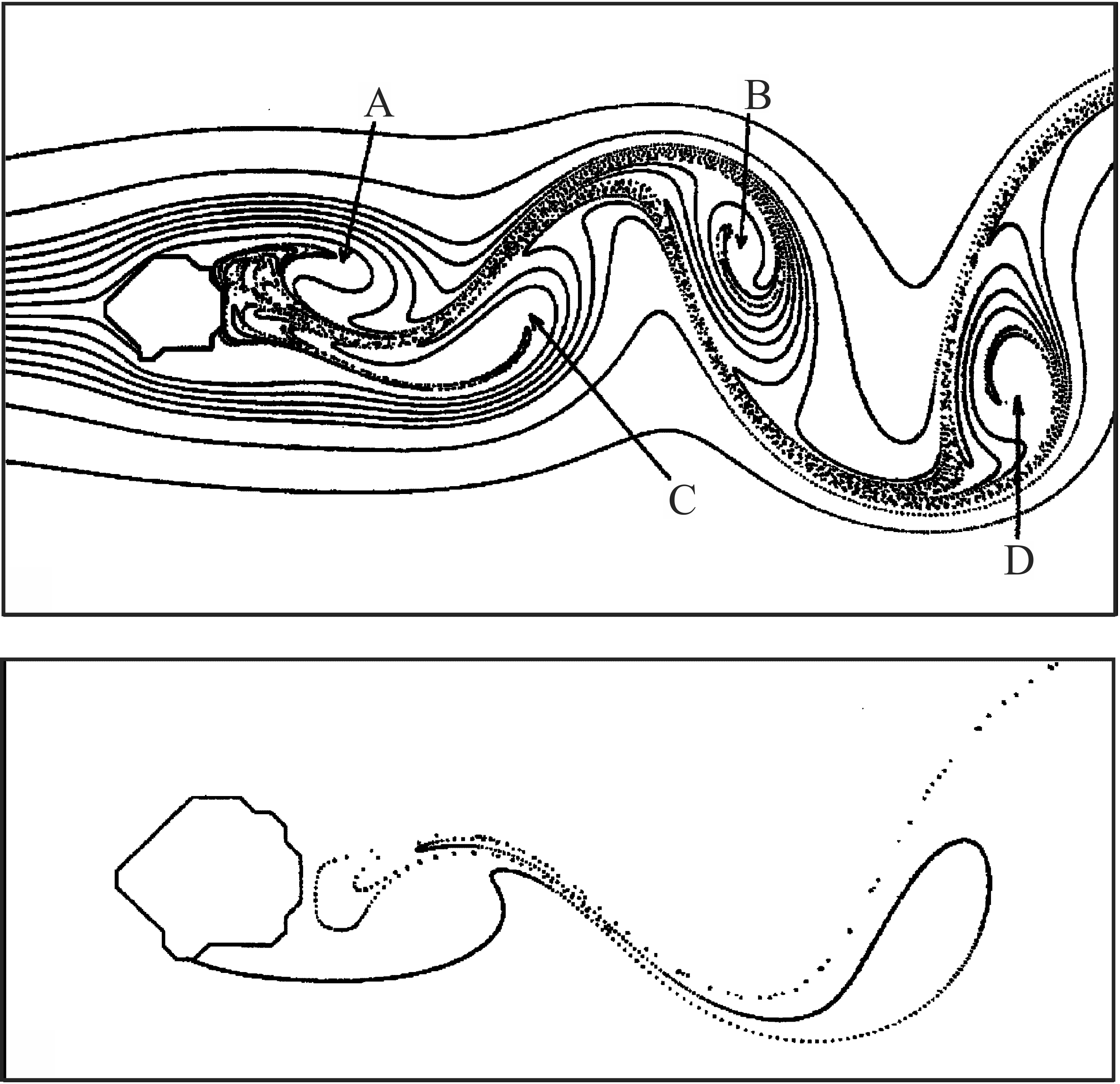}&
\includegraphics[width=0.4\columnwidth]{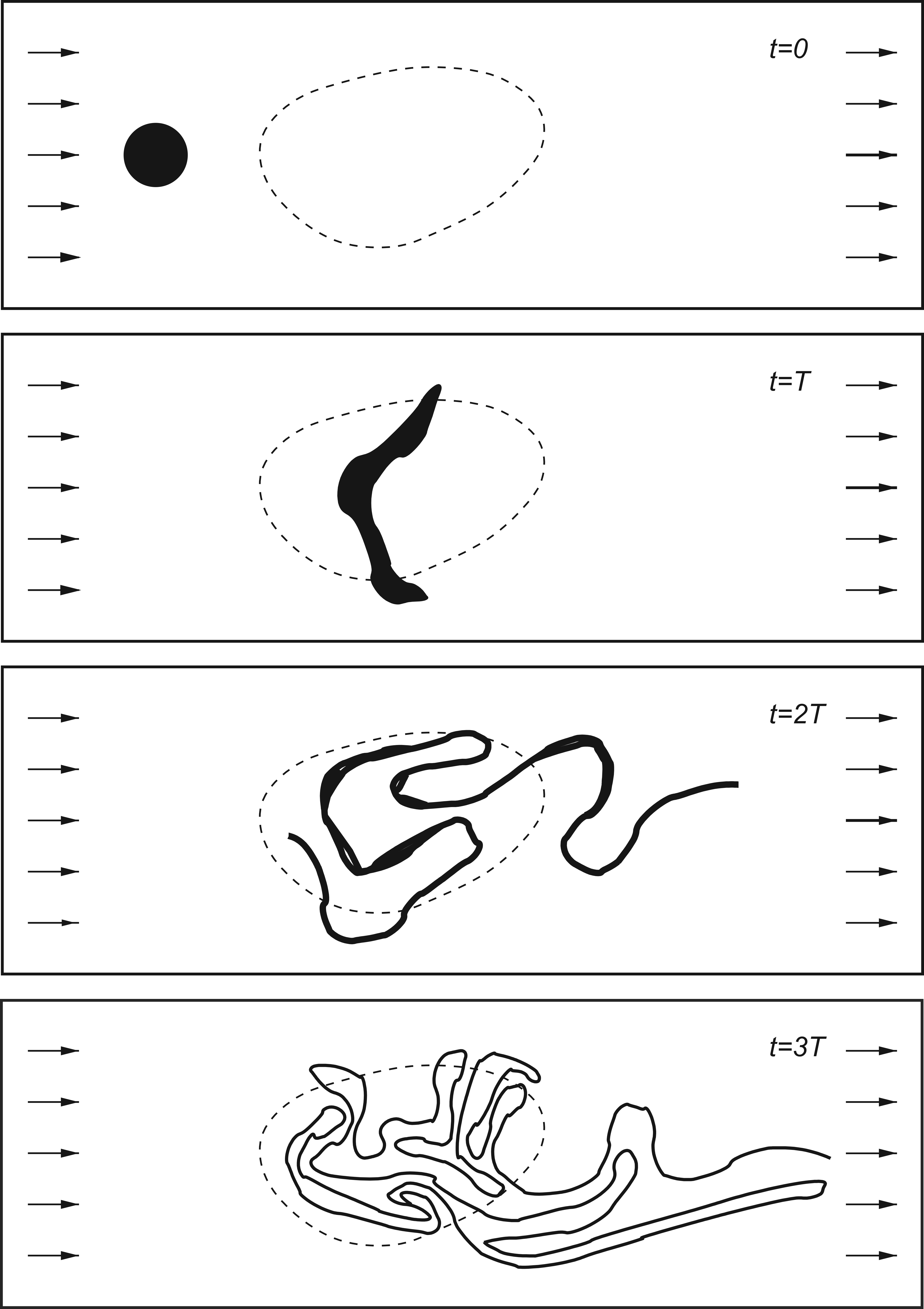}\\
(a) & (b)\\
\multicolumn{2}{c}{\includegraphics[width=0.98 \linewidth]{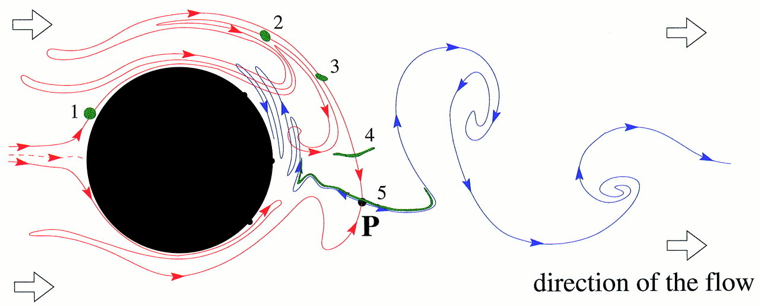}}\\
\multicolumn{2}{c}{(c)}
\end{tabular}
\end{centering}
\caption{(a) (top) Numerical simulations of particle trajectories within a von K\'arm\'an vortex street emanating from Guadalupe Island. Flow reversal due to vortices corresponds to regions indicated by A, B, C, D. (bottom) Some particle orbits have long residence times due to flow reversal (adapted from~\cite{Aristegui:1997aa}). (b) Schematic depicting trapping of of some fluid tracer particles within a mixing region (depicted by dotted line) in an open flow with increasing number of flow periods $T$ (adapted from~\cite{Tel2005}). (c) Schematic of the evolution of a blob (green) of tracer particles with number of flow periods and the associated stable (red) and unstable (blue) manifolds in the vortex-shedding wake of cylinder. Particles at the periodic point $\mathbf{P}$ remain trapped there indefinitely (adapted from~\cite{Karolyi:2000aa}).}
\label{fig:VonKarman}
\end{figure}

The interplay of transient flow reversal (as discussed in Section~\ref{subsec:hydraulic_chars}) and open flows is well-studied in fluid mechanics more broadly.  Transient flows that are relatively simple in the Eulerian frame can lead to complicated kinematics in the Lagrangian frame, as observed via particle tracking experiments and computations.   Complex Lagrangian kinematics can significantly impact transport, mixing, and reactions.  A classical example of an open flow with transient flow reversal is the von K\'arm\'an vortex street which arises from periodic vortex shedding over a bluff body as shown in Figure~\ref{fig:VonKarman}(a), (c). Here, transient flow reversal within the vortices results in the trapping of select fluid tracer particles in the wake of the flow, even though the net open flow continually passes through the flow domain. Such trapping of particles is associated with the \emph{mixing region} of the flow, a region of intense local mixing in which some fluid elements remain for arbitrarily long times (\ref{fig:VonKarman}(b)).

The von K\'arm\'an vortex street arises from the Navier-Stokes
equations that describe flow in unconfined domains, and so has
different origins and dynamics from those of the Darcy
(\ref{eq:Darcyq}) and groundwater flow equations (\ref{eq:gfe}) for
flow in porous media.  Whilst these flows are generated by different
\emph{dynamics}, the transport and mixing properties of these flows
depend only on the Lagrangian \emph{kinematics}, i.e.\/ the mapping
from the Eulerian velocity field to Lagrangian particle trajectories
as quantified by the advection equation (\ref{eq:pathline}).  As the
von K\'arm\'an and coastal aquifer flows share key structural Eulerian
characteristics (namely flow reversal in an open flow), we expect that
some of the characteristics of the Lagrangian kinematics of these
flows are similar. \cite{Tel2005} and \cite{Karolyi:2000aa} describe
the Lagrangian dynamics that lead to such trapping and mixing in open
flows, which we briefly review as follows.

As shown in Figure~\ref{fig:VonKarman}(b, c), the \emph{mixing region}
of an open flow contains the intersection of the \emph{stable} and
\emph{unstable manifolds} of the flow.  The stable manifold is a
temporally periodic (due to periodicity of the flow) material line
which is comprised of fluid tracer particles which approach the mixing
region and get ``trapped'' within this region for arbitrarily long
times ($t\rightarrow\infty$).  Whilst this appears to violate
conservation of mass, this manifold has zero area, and so particles
have a zero probability of lying on it.  Despite this, the stable
manifold has profound impacts upon transport as the residence time of
particles near the stable manifold diverges to infinity the closer a
particle is to the stable manifold.  Similar to the stable manifold,
the unstable manifold (shown in Figure~\ref{fig:VonKarman}(c)) is a
material line comprised of fluid particles which approach the mixing
region as time goes backward, and so get trapped within this region in
the limit $t\rightarrow -\infty$.  As illustrated in
Figure~\ref{fig:VonKarman}(b), fluid tracer particles close to the
unstable manifold eventually leave the mixing region, and so the
unstable manifold can be directly observed in open flows by placing a
``blob'' of fluid particles over the stable manifold.  Once this blob
enters the mixing region, many particles are swept downstream out of
the mixing region; the remainder are trapped locally near the unstable
manifold.  The remaining particles continue to leak out with time,
forming a locus that closely approximates the unstable manifold.

The particles which remain trapped in the mixing region lie close to
the \emph{chaotic saddle}, a fractal set of points which never leave
the mixing region. As shown in Figure~\ref{fig:VonKarman}(c), the
points which comprise the chaotic saddle are formed by the
intersections of the stable and unstable manifolds. These
intersections generate continual stretching and folding of fluid
elements ~\citep[the hallmark of chaotic dynamics,][]{Ottino:1989aa}
in the chaotic saddle, leading to chaotic advection and exponential
stretching of fluid elements within the mixing region.

It is the interplay of flow reversal and open flow that generates the complex
mixing and transport dynamics that cannot exist in steady or isotropic
Darcy flow.  Specifically, the mixing regions of such reversing flows
generate chaotic advection which leads to rapid mixing and
homogenisation, whilst the chaotic saddle traps neighboring fluid
particles for arbitrarily long times.  In addition, the mixing region
can also admit non-mixing ``islands'' which (in contrast to the
chaotic saddle) are of non-zero area, and so can trap finite amounts
of fluid for infinitely long times.  All of these dynamics lead to
strongly anomalous transport.  Conversely, steady Darcy flows are
topologically simple (i.e. they have well-defined streamlines and
streamsurfaces) and cannot have stagnation points, hence flow
reversal in periodic discharge systems is a marked departure
from  mixing and transport in steady systems.  In the remainder of this paper we
seek to understand the prevalence of flow reversal in tidally forced
aquifers and the impacts on mixing and transport.  Using the
numerical model described in Section~\ref{sec:model} we study
the Lagrangian kinematics of these flows and consider the implications
for anomalous transport.

\section{Non-dimensionalization of the governing equations}

In order to address the study of Lagrangian kinematics we need to generate accurate solutions to tidally forced groundwater systems. In principle any groundwater flow package can be used to solve the governing equations (\ref{eq:Helmholtz})--(\ref{eq:nvshperiodic}), but extra care must be taken to assure that accurately cyclo-stationary solutions satisfy the periodic equation in (\ref{eq:Helmholtz}). The finite-difference algorithm of \citet{Trefry:2009aa} is second-order convergent and efficiently provides direct solutions (i.e. without time stepping) for both steady and periodic equations. Input parameters to the solution algorithm are the $K$ distribution, the imposed regional flux gradient $J$, the tidal amplitude $\textsl{g}_{p}$, the modal frequency $\omega$, the storativity $S$, and the reference porosity $\varphi_{\textrm{ref}}$. It is useful to non-dimensionalize the governing equations and identify the key dimensionless parameters, which are summarised as follows.

\subsection{Dimensionless parameters}

\subsubsection{Heterogeneity model}
The choice of $K$ distribution deserves some discussion. Characterization and representation of physically realistic conductivity fields is a non-trivial task and since the early work of \citet{Delhomme:1979aa} a variety of geostatistical inversion techniques have been developed to make best use of the often sparse field measurements \citep[see for example][]{DeutschJournel:1992aa,Ezzedine:1999aa,Fienen:2009aa}. In contrast, in the present work we seek to identify transferable attributes of tidally forced flows, so our emphasis is on understanding how simple models of spatial heterogeneity may potentially contribute to enhanced mixing processes.  Our expectation is that if enhanced mixing is detected in simple fields, then similar mixing dynamics will likely also appear in more sophisticated, better-conditioned (and more representative) geostatistical fields. Thus we restrict attention to random spatial processes governed by Gaussian autocorrelation functions, where the mean ($K_{\textrm{eff}}$), log-variance ($\sigma^{2}_{\textrm{log}K}$) and integral scale ($\lambda$) are sufficient to describe the statistics.

\subsubsection{Townley number $\mathcal{T}$}
\citet{Townley:1995aa} shows that the dimensionless \emph{Townley number} $\mathcal{T}$ captures the relative timescales of diffusion $\tau_D=L^2/D_{\textrm{eff}}$ and tidal forcing $\tau_T=1/\omega$ in a finite tidal aquifer as
\begin{equation} \label{eq:TownleyNum}
\mathcal{T} \equiv \frac{\tau_D}{\tau_T}=\dfrac{L^2 S \omega}{K_{\textrm{eff}}} = \dfrac{2 \pi L^2}{D_{\textrm{eff}} P}
\end{equation}
where $D_{\textrm{eff}} \equiv K_{\textrm{eff}}/S$ is the effective aquifer diffusivity. As is well understood for homogeneous aquifers (see following sections), low values of $\mathcal{T}$ provide conditions conducive to propagation of tidal signals far into the aquifer with low phase lags, while high $\mathcal{T}$ values ensure rapid attenuation (damping) of tidal amplitudes and phase lags growing rapidly with increasing penetration distance. Here, for convenience, we are interested in systems where $\mathcal{T}$ is large enough to ensure that finite-aquifer effects are negligible in the tidal forcing zone. 

\subsubsection{Tidal strength $\mathcal{G}$}
We also characterise the relative strength of the tidal forcing amplitude $\textsl{g}_p$ to the inland regional gradient $J$, thereby defining the \emph{tidal strength}
\begin{equation} \label{eq:Tidalstrength}
\mathcal{G} \equiv \frac{\textsl{g}_p}{J L},
\end{equation}
such that $\mathcal{G} = 0$ corresponds to conventional steady
discharge to a constant fixed head boundary.  From
(\ref{eq:Helmholtz}), (\ref{eq:hmbc})
$\mathcal{G}=\langle||\mathbf{q}_{p}||\rangle/\langle||\mathbf{q}_{s}||\rangle$
in the limit of slow forcing $\omega\rightarrow\infty$, where the
angle brackets denote a spatial average over the flow domain.  Hence
the tidal strength $\mathcal{G}$ controls the propensity for flow
reversal over a forcing cycle.  Indeed, it can be shown numerically
that the product of tidal strength and aquifer log-conductivity
variance
\begin{equation} \label{eq:FlowReversal}
\mathcal{G} \, \sigma^2_{\log K}=\frac{g_\text{p}}{J L}\,\sigma^2_{\log K},
\end{equation}
is strongly correlated with the density of canonical flux ellipses in the tidally active zone, where $\mathcal{G}\, \sigma^2_{\log K} = 0$ corresponds to flow in aquifers with homogeneous hydraulic conductivity or zero tidal forcing.

\subsubsection{Tidal compression ratio $\mathcal{C}$}
The final main dimensionless parameter is the tidal compression ratio which characterises the relative change in porosity of the aquifer from its reference state ($\varphi_{\textrm{ref}}$, $h_{\textrm{ref}}$) under a pressure fluctuation of the same magnitude as the tidal forcing amplitude ($\Delta h = \textsl{g}_p$), i.e.
\begin{equation} \label{eq:Aquifercompressibility}
\mathcal{C} \equiv \frac{\varphi-\varphi_{\textrm{ref}}}{\varphi_{\textrm{ref}}} = \frac{S \textsl{g}_p}{\varphi_{\textrm{ref}}},
\end{equation}
such that $\mathcal{C}=0$ corresponds to an incompressible aquifer and
$\mathcal{C} \ll 1$ corresponds to a weakly compressible aquifer. These
limits shall prove useful in understanding how chaotic mixing arises
in strongly compressible aquifers, i.e. $\mathcal{C} \lesssim 1$.

The dimensionless parameter set $\mathcal{Q}\equiv(\mathcal{T},\mathcal{G},\mathcal{C})$ controls the dynamics of a periodically forced tidal aquifer. In combination with the set of statistical parameters $\chi$ which define the dimensionless hydraulic conductivity field (where for the log-Gaussian conductivity field in this study $\chi=(\sigma^2_{\ln K},\lambda)$), this parameter set completely defines the dimensionless transient tidal forcing problem and so these parameters serve as model inputs. In this way, the set of dynamical parameters $\mathcal{Q}$ can then be translated between aquifer models with different conductivity structures (as defined by $\chi$). 

\subsection{Heterogeneity characters $\mathcal{H}_t$ and $\mathcal{H}_x$}
In addition to the input parameters sets $\mathcal{Q}$ and $\chi$ there also exist two \emph{characteristic}
dimensionless parameters, $\mathcal{H}_t$ and $\mathcal{H}_x$, that are functions of the input parameters. These heterogeneity characters are completely defined by $\mathcal{Q}$ and $\chi$ and aid understanding of how the aquifer transport dynamics relate to the heterogeneous structure of the conductivity field.

The heterogeneous flow is governed by the interaction of two physical processes with independent time scales. First, through the imposed regional gradient the inland part of the domain displays a mean \emph{drift velocity} $v_{\textrm{drift}} \equiv -K_{\textrm{eff}} J/\varphi_{\textrm{ref}}$ toward the tidal boundary where the fluid ultimately discharges. Fluid parcels advecting within this mean drift sample successive $K$ heterogeneities on a time scale of $t_{\textrm{drift}} = \lambda/|v_{\textrm{drift}}| = \lambda \varphi_{\textrm{ref}}/K_{\textrm{eff}} |J|$. Second, the tidal boundary oscillates with period $P$. We define the \emph{temporal character} of the heterogeneous flow, $\mathcal{H}_{t}$, as the ratio of the drift time scale to the tidal period, i.e. $\mathcal{H}_{t} \equiv t_{\textrm{drift}}/P = t_{\textrm{drift}} \omega/2\pi$. When $\mathcal{H}_{t} \ll 1$ the system is said to be \emph{discharge dominated} and the groundwater flow displays minimal lateral (longshore) deflections and residence time variances scale with $\sigma^{2}_{\textrm{log}K}$. For $\mathcal{H}_{t} \gg 1$ the system is \emph{tidally dominated} with low drift velocity: although fluid parcels experience many tidal periods during the journey to the discharge boundary, lateral deflections of the flow paths are suppressed due to the low velocity. Where $\mathcal{H}_{t} \approx \mathcal{O}(1)$ the system is in \emph{temporal resonance} and there is maximum potential for local elliptical velocity orbits to induce folding of flow paths and the development of chaotic structures.

We also introduce the \emph{spatial character} of the heterogeneous flow, $\mathcal{H}_{x}$, as a measure of the density of heterogeneities in the tidally affected zone. We define the \emph{tidally affected zone} as the zone from the tidal boundary to the interior point, $x_{taz}$, where the amplitude of the tidal oscillation ($g_{p}$) matches the mean local steady head, i.e. $|h_{p}(x_{\textrm{taz}},t)| = h_{s}(x_{\textrm{taz}})$. $h_{p}$ can conveniently be estimated using a one-dimensional homogeneous model \citep{Townley:1995aa}, fixing $x_{\textrm{taz}}$ independently of the heterogeneous numerical solution. The relevant analytical solutions are
\begin{equation} \label{eq:tazup}
h_{s}(x) = x J L \: ; \: h_{p}(x,t) = \textsl{g}_{p}\: \textrm{cosh}\left[(x-1)\sqrt{i \: \mathcal{T}}\right] \: \textrm{sech}\left[\sqrt{i \: \mathcal{T}}\right] e^{i \omega t},
\end{equation}
which are easily established by integration \citep{Trefry:2011aa}. It is straightforward to show that $x_{\textrm{taz}}$ is the root of
\begin{equation}
\label{eq:tazroot}
\mathcal{G} \sqrt{\frac{\cos \left((x_{\textrm{taz}}-1) b \right)+\cosh \left((x_{\textrm{taz}}-1) b \right)}{\cos \left(b \right)+\cosh \left(b \right)}} - x_{\textrm{taz}} = 0,
\end{equation}
where $b = \sqrt{2 \mathcal{T}}$.  We define the spatial character by
$\mathcal{H}_{x} \equiv x_{taz}/\lambda$, which expresses the number
of spatial correlation scales of $K$ that fit within the width of the
tidally affected zone (perpendicular to the boundary).  The higher
$\mathcal{H}_{x}$, the greater the number of conductivity contrasts
encountered by the discharging flow while subject to strong elliptical
motions.

\subsection{Scaled equations and solution approach} \label{sec:model}

\subsubsection{Dimensionless model}

Based on these physical parameters, we write the governing equations (\ref{eq:Helmholtz})-(\ref{eq:hmbc}) in dimensionless form via the rescalings $\mathbf{x}' = (x',y') = \mathbf{x}/L = (x/L,y/L)$, $h'=h/(JL)$, $t'=t\omega$, $\kappa=K/K_{\textrm{eff}}$ yielding the non-dimensional governing equations (where primes are henceforth dropped)
\begin{align} \label{eq:Helmholtz_nondim}
\nabla\cdot(\kappa\nabla h_{s}(\mathbf{x})) = 0, & & \nabla\cdot(\kappa\nabla h_{p}(\mathbf{x})) - i \mathcal{T} h_{p}(\mathbf{x}) = 0,
\end{align}
and non-dimensional boundary conditions on $\mathcal{D}$ (now the unit square)
\begin{align} \label{eq:hmbc_nondim}
h_{s}(1,y) = 1, & & h_{s}(0,y) =0, & & h_{p}(1,y) = 0, & & h_{p}(0,y)& =\mathcal{G}.
\end{align}
The parameter set $(\mathcal{T},\mathcal{G})$ governs the dimensionless Darcy flux $\mathbf{q}'=\mathbf{q}/(JK_{\textrm{eff}})$. Conversely, the tidal compression ratio $\mathcal{C}$ governs scaling of the dimensionless velocity as
\begin{equation}
\mathbf{v}'=\mathbf{v}\frac{\varphi_{\textrm{ref}}}{JK_{\textrm{eff}}}=\mathcal{C}\mathbf{q}'.
\label{eqn:vel_dless}
\end{equation}
The Lagrangian kinematics are governed by the dynamical parameter set $\mathcal{Q}\equiv(\mathcal{T},\mathcal{G}, \mathcal{C})$.

\subsubsection{Lagrangian particle tracking}

Given solution of (\ref{eq:Helmholtz_nondim}), (\ref{eq:hmbc_nondim})
via the finite difference (FD) method described in
\cite{Trefry:2011aa}, the Darcy flux $\mathbf{q}$ and fluid velocity
$\mathbf{v}$ are computed via (\ref{eq:Darcyq}) and
(\ref{eq:vsingle}), respectively. To probe the Lagrangian kinematics
of these flows, we integrate the advection equation
(\ref{eq:pathline}) for many passive fluid tracers over many millions
of periods of the flow.  As shall be shown, in conjunction with the
tools and techniques of \emph{measure-preserving} dynamical systems
(chaos theory), such analysis in the \emph{Lagrangian frame} allows a
clear visualisation of the transport dynamics that may not be
otherwise apparent.

We require interpolated values of the Darcy flux and fluid velocity
for locations away from the FD grid that exactly satisfy
(\ref{eq:Helmholtz_nondim}), as even minor errors violate the
measure-preserving nature of the system and lead to spurious results
(discussed below).  Whilst the FD method computes
(\ref{eq:Helmholtz_nondim}) to within machine precision with respect
to the FD stencil (which approximates the differential operators in
(\ref{eq:Helmholtz_nondim})), interpolated (off-grid) values of
$\mathbf{q}_s(\mathbf{x},t)$ and $\mathbf{q}_p(\mathbf{x},t)$ do not
exactly satisfy the continuous differential operators of the governing
equations.  From the continuity equation (\ref{eq:cedeform}), the
steady Darcy flux $\mathbf{q}_s(\mathbf{x})$ must be exactly
divergence free, and, as described in Appendix~\ref{app:numerical_ce},
we developed a spline interpolation streamfunction representation of
$\mathbf{q}_s(\mathbf{x})$.  The periodic flux
$\mathbf{q}_p(\mathbf{x},t)$ also must individually satisfy the
continuity equation (\ref{eq:cedeform}); this is achieved by
interpolating $\mathbf{q}_p(\mathbf{x})$ from the FD grid and then
computing the periodic contribution to the porosity $\varphi_p$ from
the divergence of $\mathbf{q}_p(\mathbf{x})$, also described in
Appendix~\ref{app:numerical_ce}.  By constructing $\textbf{q}$ and
$\varphi$ in this manner we ensure the solution obeys the continuity
equation exactly and the subsequent velocity $\mathbf{v}$ conserves
mass to machine precision at \emph{all} interior locations.  This is
absolutely critical for the long-term integration of particle
trajectories (via (\ref{eq:pathline})) and analysis of Lagrangian
characteristics because even tiny violations of mass conservation can
lead to serious errors such as spurious sources and sinks in the flow
domain and violation of Lagrangian topology~\citep{Ravu:2016aa}.

\section{Fundamentals of tidal flows in heterogeneous domains}

\begin{figure}
\begin{centering}
\begin{tabular}{ccc}
\includegraphics[width=0.39\columnwidth]{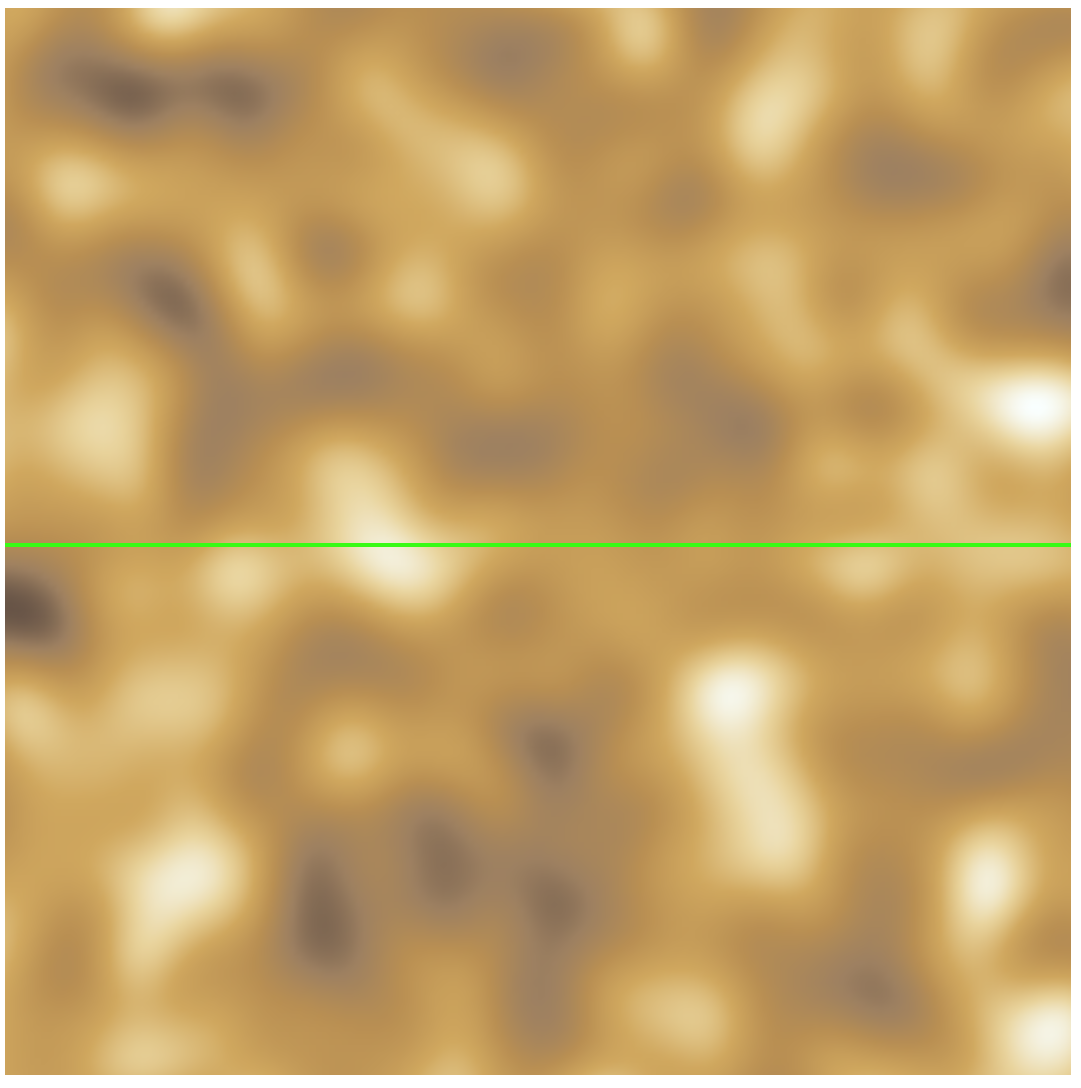}&
\includegraphics[width=0.42\columnwidth]{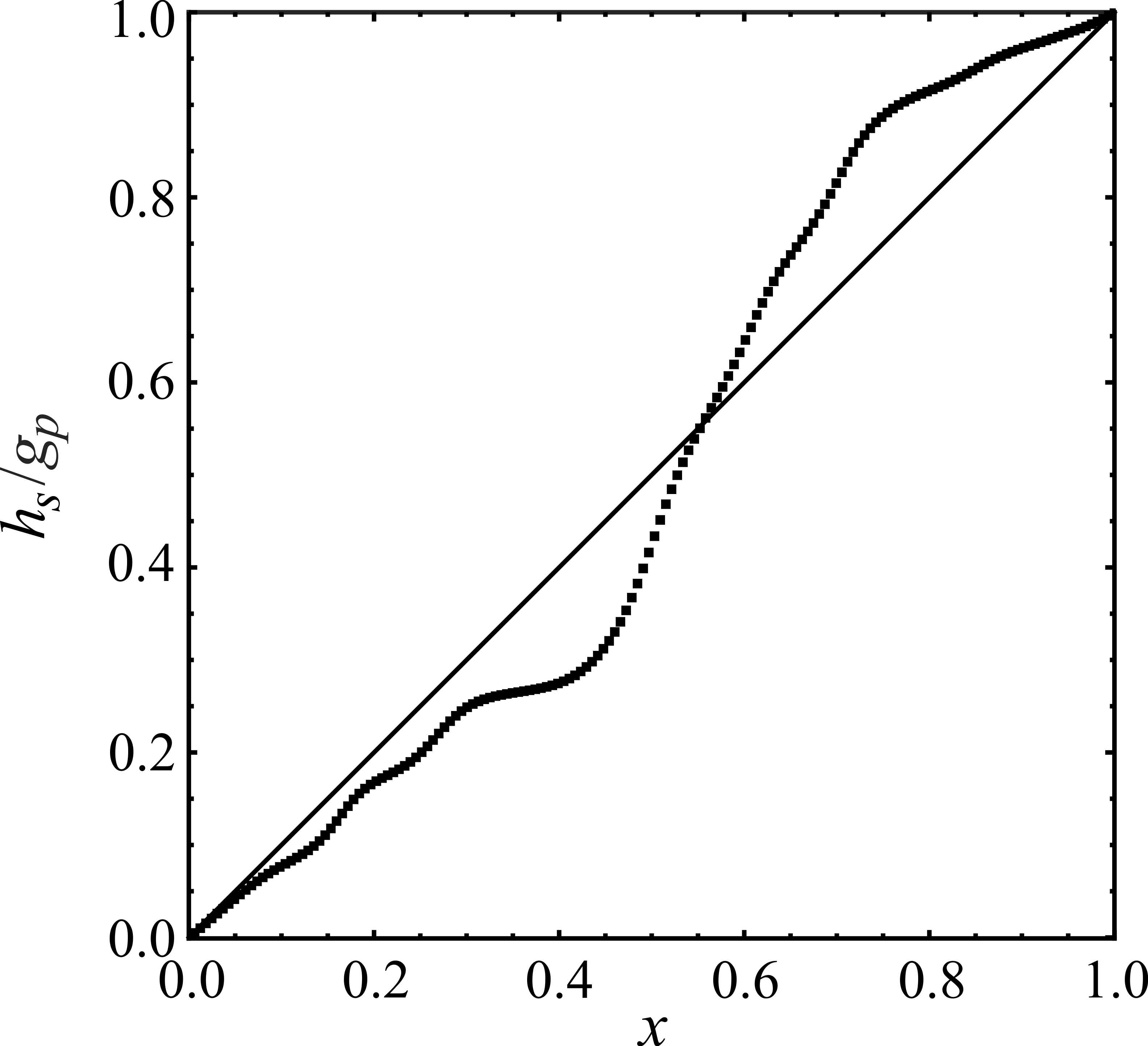}\\
(a) & (b)\\
\includegraphics[width=0.445\columnwidth]{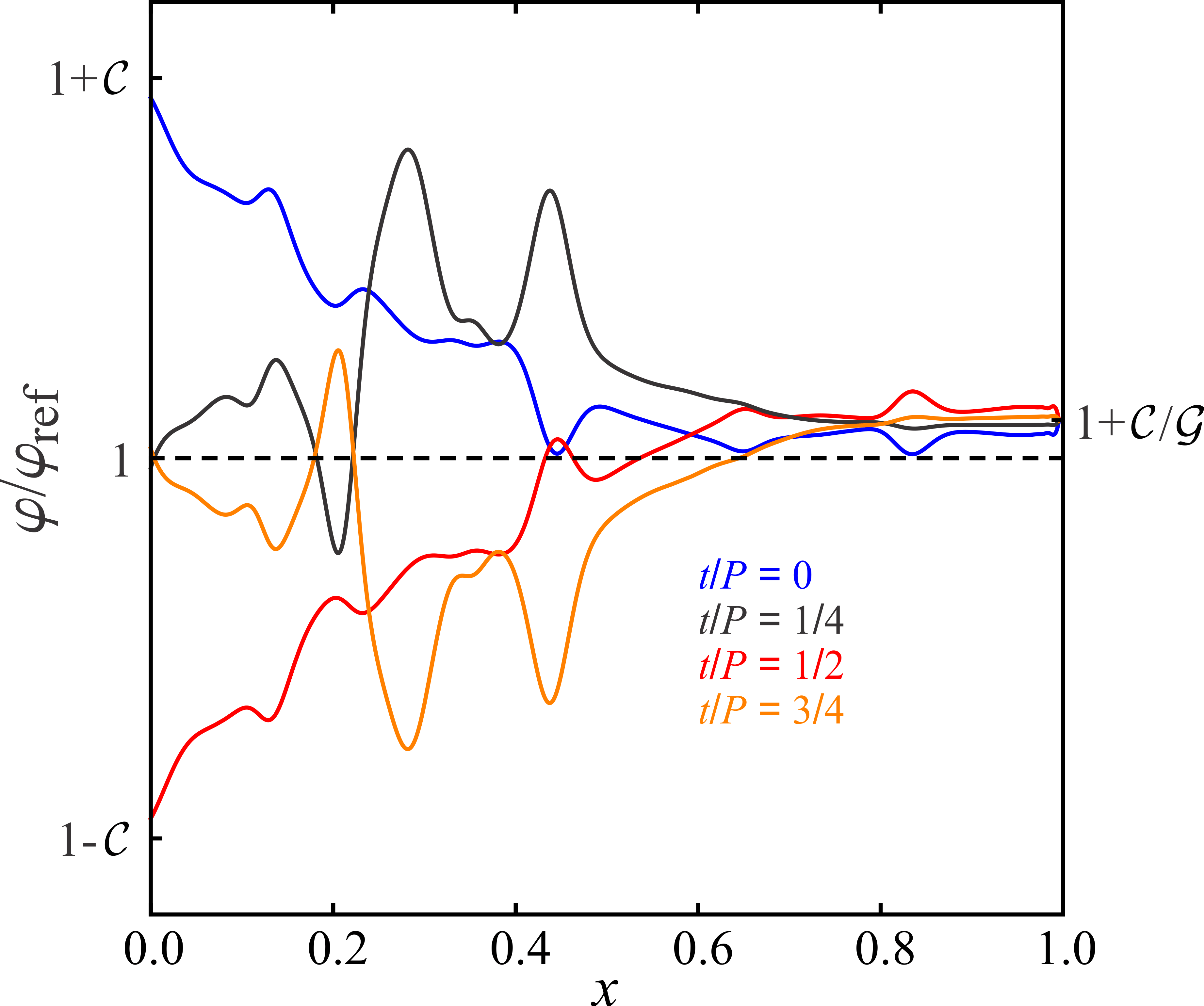}&\includegraphics[width=0.42\columnwidth]{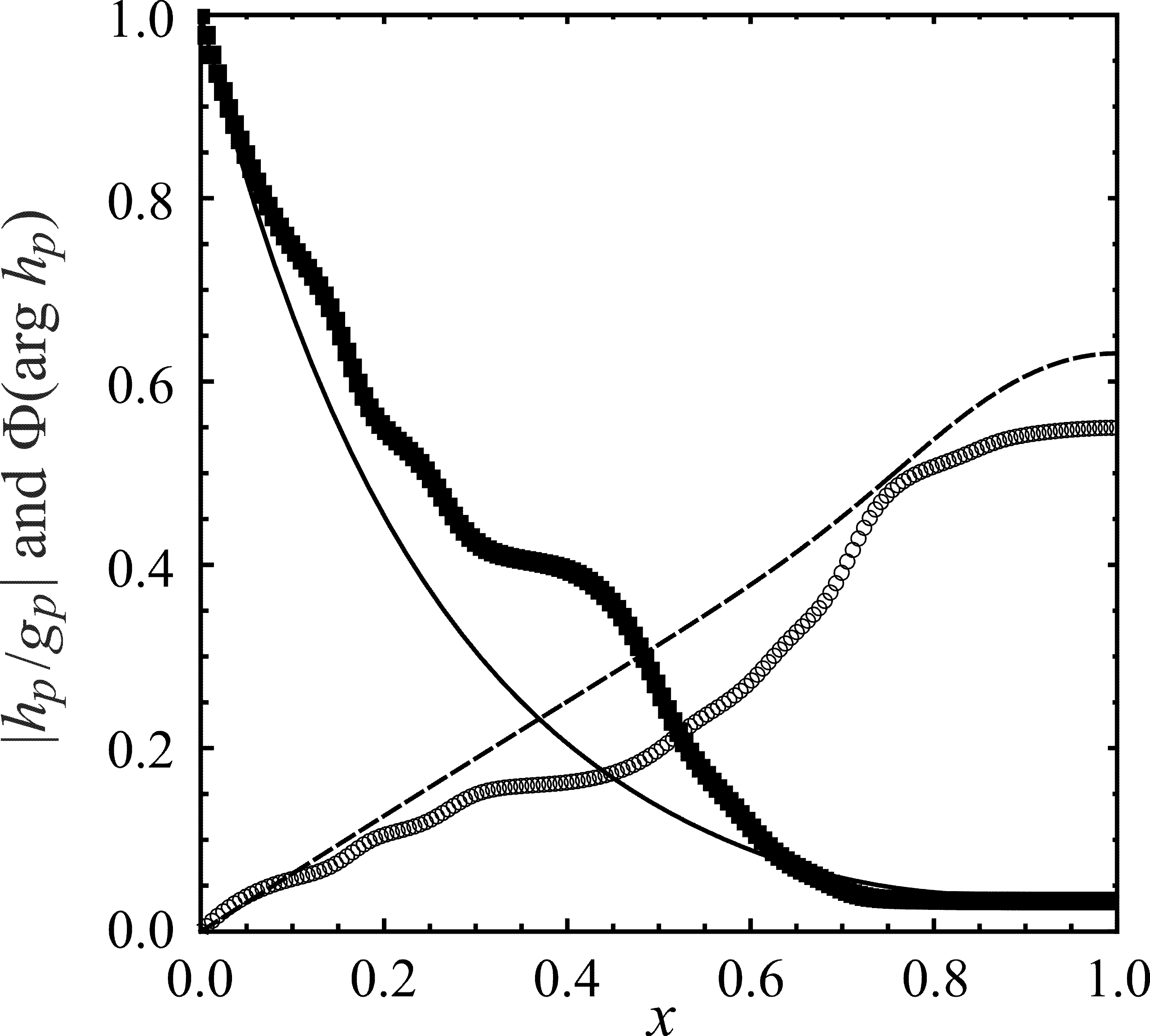}\\
(c) & (d)
\end{tabular}
\end{centering}
\caption{(a) Density map of log hydraulic conductivity $\kappa$ where
  lower (higher) values are darker (lighter).  (b) Profile of the
  steady head $h_s$ for the heterogeneous simulation (squares) and
  reference homogeneous solution (line) along the green line $y$=0.5
  shown in (a).  (c) Variation of the normalized total porosity
  $\varphi/\varphi_{ref}$ along $y$=0.5 over the tidal cycle.  (d)
  Same as for (b) but showing the periodic head solution $h_{p}$
  (square) and phase lag $\arg h_{p}$ (circles) for the heterogeneous
  simulation with the reference homogeneous solutions shown as
  curves.}
\label{fig:kappadist}
\end{figure}

In this study we focus on the Lagrangian kinematics of the tidally
forced groundwater system described in Section~\ref{sec:model} with
parameter set $(\mathcal{T},\mathcal{G},\mathcal{C})$ = (10$\pi$, 10,
0.5) which corresponds to a highly compressible and diffusive aquifer
system subject to a diurnal tidal signal.  In Section~\ref{sec:field}
we place this parameter set in context with field studies.  In
subsequent studies we shall explore the distribution of Lagrangian
kinematics over the coastal aquifer parameter space
$\mathcal{Q}=\mathcal{T}\times\mathcal{G}\times\mathcal{C}$.  The
hydraulic conductivity field $\kappa$ used is a single
$(164 \times 164)$ realization generated according to the algorithm of
\citet{Ruan:1998aa} corresponding to an aquifer of moderate
heterogeneity
($\lambda = 0.049, \sigma^{2}_{\textrm{log}\kappa} = 2$).  This field
yields reversal number $\mathcal{G}\, \sigma^2_{\log K}=20$ and
heterogeneity characters $(\mathcal{H}_t,\mathcal{H}_x) = (4.9,13.6)$,
i.e. the system is near temporal resonance and has $\mathcal{O}(10)$
correlation scales within the tidally active zone.  The $\kappa$ field
is shown in Figure~\ref{fig:kappadist}, along with the associated
steady and periodic head components and evolution of the porosity
profile over a forcing cycle.  These plots clearly show an exponential
decay oscillation amplitude with distance from the tidal boundary for
both heterogeneous and homogeneous aquifers, along with a phase lag
that increases with distance.  Fluctuations in the heterogeneous head
solutions away from the homogeneous state have spatial scales that are
somewhat larger than $\lambda$ \citep{Trefry:2011aa}.  As shown in
Figure~\ref{fig:kappadist}d, the normalized porosity
($\varphi/\varphi_{\textrm{ref}}$) oscillates around unity (with
amplitude $\mathcal{C}$) synchronously with the tidal condition at the
left boundary, and tends to a value determined by the upstream
boundary head
$\varphi/\varphi_{\textrm{ref}} \rightarrow 1 + J L
S/\varphi_{\textrm{ref}} = 1 + \mathcal{C}/\mathcal{G}$ as
$x \rightarrow 1$.

\begin{figure}[p]
\begin{centering}
\begin{tabular}{c}
\includegraphics[width=0.92\columnwidth]{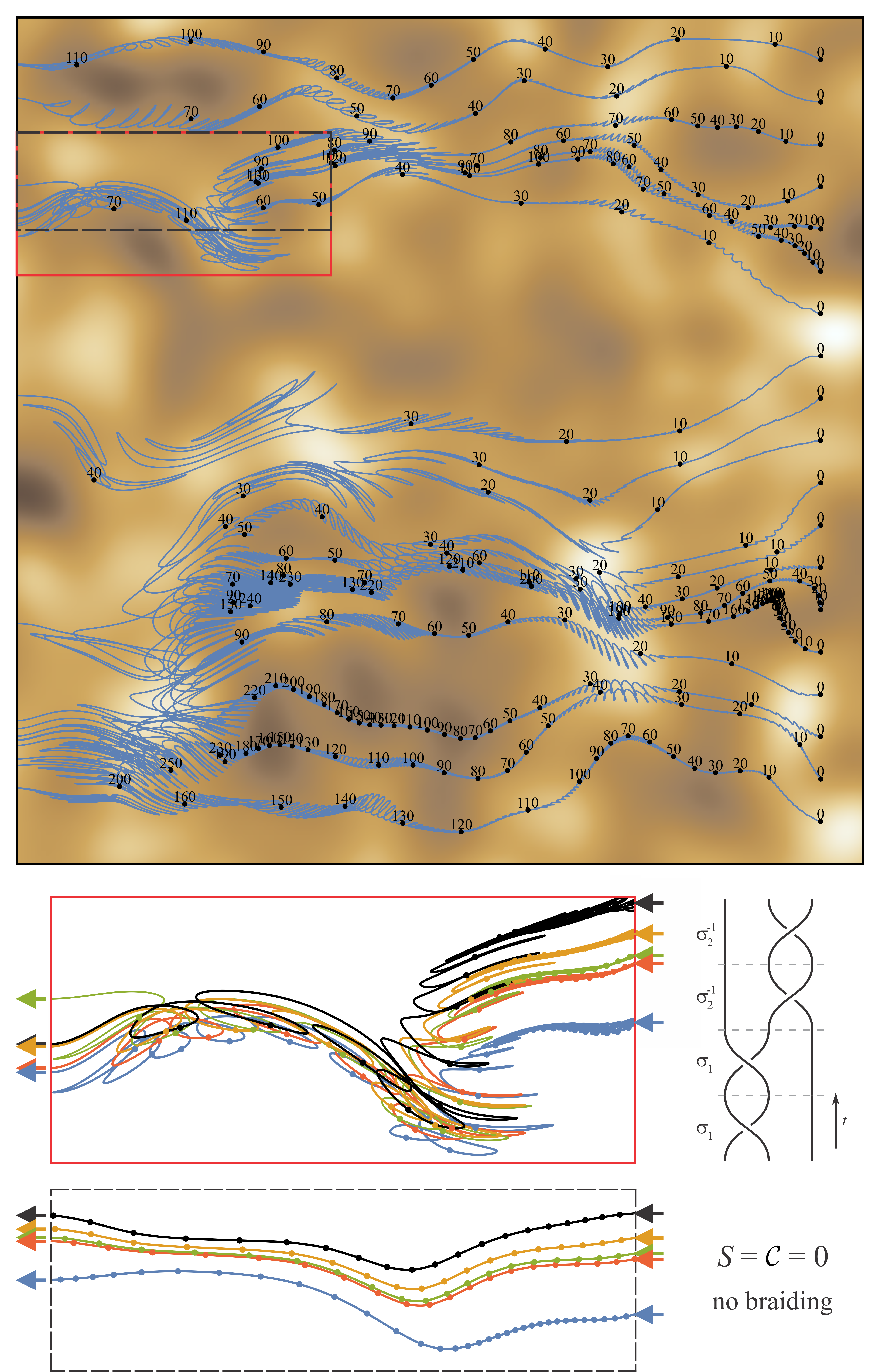}
\end{tabular}
\end{centering}
\caption{(top) Flow paths, with the number of flow periods shown along each trajectory. (middle left) Detail of the red box with flow paths coloured and dotted every period. Note the change in vertical order of the flow paths from entry to exit (braiding). (middle right) Schematic of generic braiding of fluid trajectories with time $t$ (adapted from~\cite{Finn:2011aa}). (bottom) Corresponding flow paths for the zero-compression case -- no braiding is possible (see Appendix \ref{app:zero_compression}).}
\label{fig:flowpaths}
\end{figure}

Of primary interest are the flow paths shown in
Figure~\ref{fig:flowpaths} which are calculated from a set of equally
spaced starting locations near the inland boundary $x=1$; the elapsed
number of flow periods is annotated along each flow path. As expected,
the spatial heterogeneity of $\kappa$ causes significant deflections
of flow paths around low conductivity zones, and travel speeds also
vary greatly along and between flow paths. In addition, as the paths
move closer to the tidal boundary they encounter flow reversal cells,
causing some paths to undergo many sweeps back and forth (flow
reversals) with relatively slow net progress toward the tidal
boundary. Such flow reversal can result in \emph{braiding} of particle
trajectories, as clearly evidenced by the different vertical ordering
of the trajectories at the left and right boundaries in
Figure~\ref{fig:flowpaths} (bottom left). Note that such braiding and
transient ``crossing'' of streamlines can be a signature of chaotic
mixing, but this is not possible in steady 2D or in incompressible
Darcy flows.

Recent studies~\citep{Finn:2011aa, Thiffeault:2010aa} have shown that
symbolic representation of the braiding motions shown in
Figure~\ref{fig:flowpaths} (lower right) allow the complexity of the
braiding motions to be calculated (in terms of the so-called
\emph{topological entropy}), which provides an accurate lower bound
for the Lyapunov exponent that characterises chaotic mixing. Trivial
braids, such as two sequential crossings that cancel each other out
have zero complexity, indicating non-chaotic kinematics.  Although not
evaluated here, the orbits shown in Figure~\ref{fig:flowpaths} (bottom
left) indicate non-trivial braids with positive complexity and hence
chaotic dynamics.

\section{Mechanisms and measures of complexity}
As seen in the previous section, flow paths for the heterogeneous tidal problem can become entwined and generally present a complicated picture that may obscure the underlying dynamical structures. In this section we seek methods to elucidate and quantify the structure of the tidal discharge map.

\subsection{Flow reversal ellipses}

\begin{figure}
\centering
\begin{tabular}{c c}
\includegraphics[width=0.57\columnwidth]{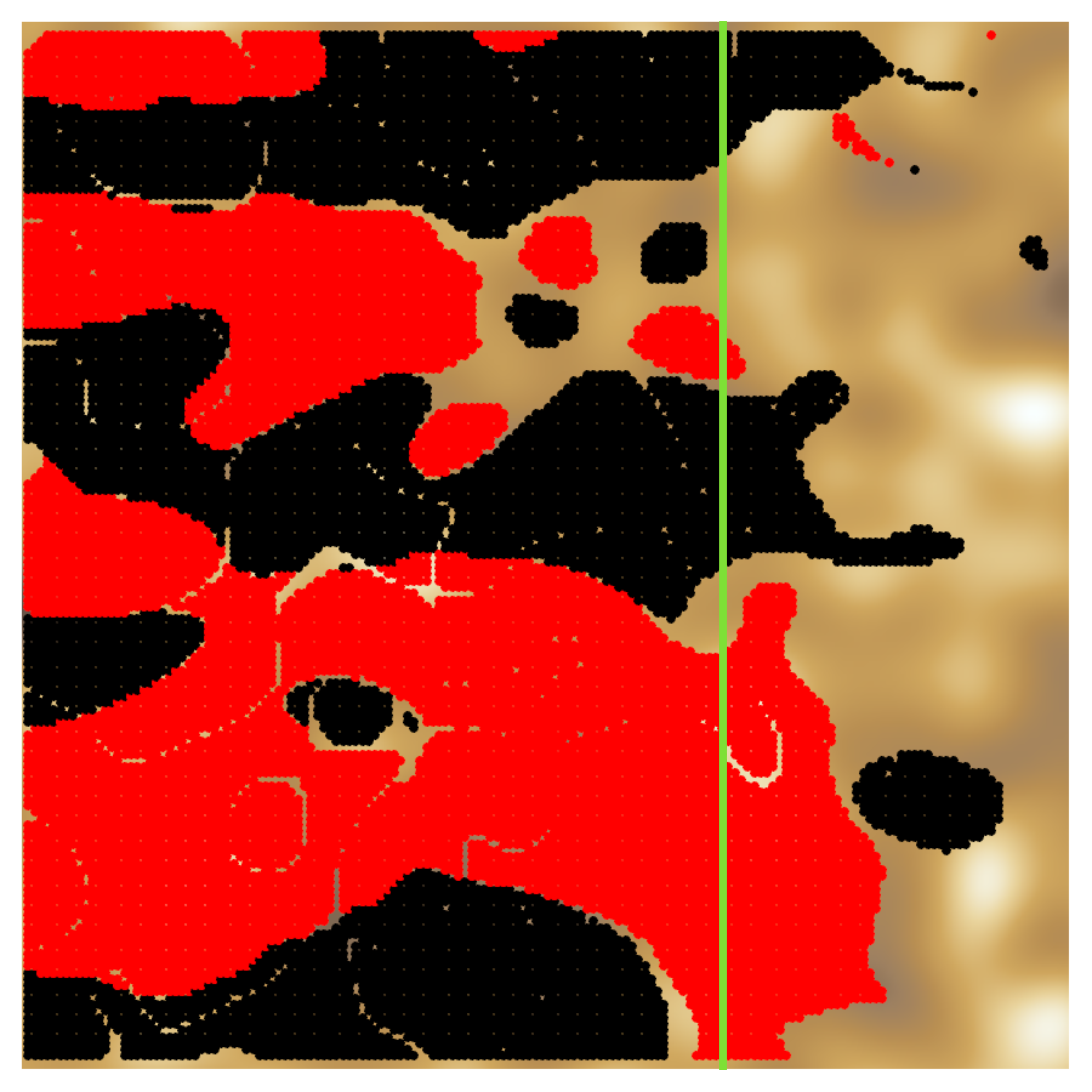}
\end{tabular}
\caption{Location of clockwise (black) and anticlockwise (red)
  canonical flux ellipses (red/black points) evaluated over the
  aquifer domain.  The edge of the canonical flux ellipse distribution
  approximates the extent of the tidally active zone shown by the
  green line drawn at $x=x_{\textrm{taz}}$.}
\label{fig:canonicalzone}
\end{figure}

Figure~\ref{fig:canonicalzone} shows that almost all of the tidally
active zone undergoes flow reversal, where the distribution of flow
reversal ellipses in the model aquifer domain is dense near the tidal
boundary, and these ellipses appear to extend into the aquifer domain
with mean distance similar to $x_{\textrm{taz}}$. The distribution of
ellipse rotation (clockwise/anti-clockwise) is also heterogeneous, and
is spatially correlated with a somewhat larger integral scale than the
underlying hydraulic conductivity field. Whilst flow reversal is
directly involved in the generation of complex Lagrangian kinematics
(as discussed in Section~\ref{chaotic_saddles}), we find little
correlation between the distribution of flow reversal ellipses and
their orientation and the Lagrangian topology of the flow which is
discussed throughout this section.  Lack of correlation between
Eulerian flow measures (e.g. flow reversal) and Lagrangian kinematics
and topology is characteristic of chaotic flows and highlights the
necessity of visualising transport in the Lagrangian frame.

\subsection{Residence time distributions}
\begin{figure}
\begin{centering}
\begin{tabular}{c}
  \includegraphics[width=0.65\columnwidth]{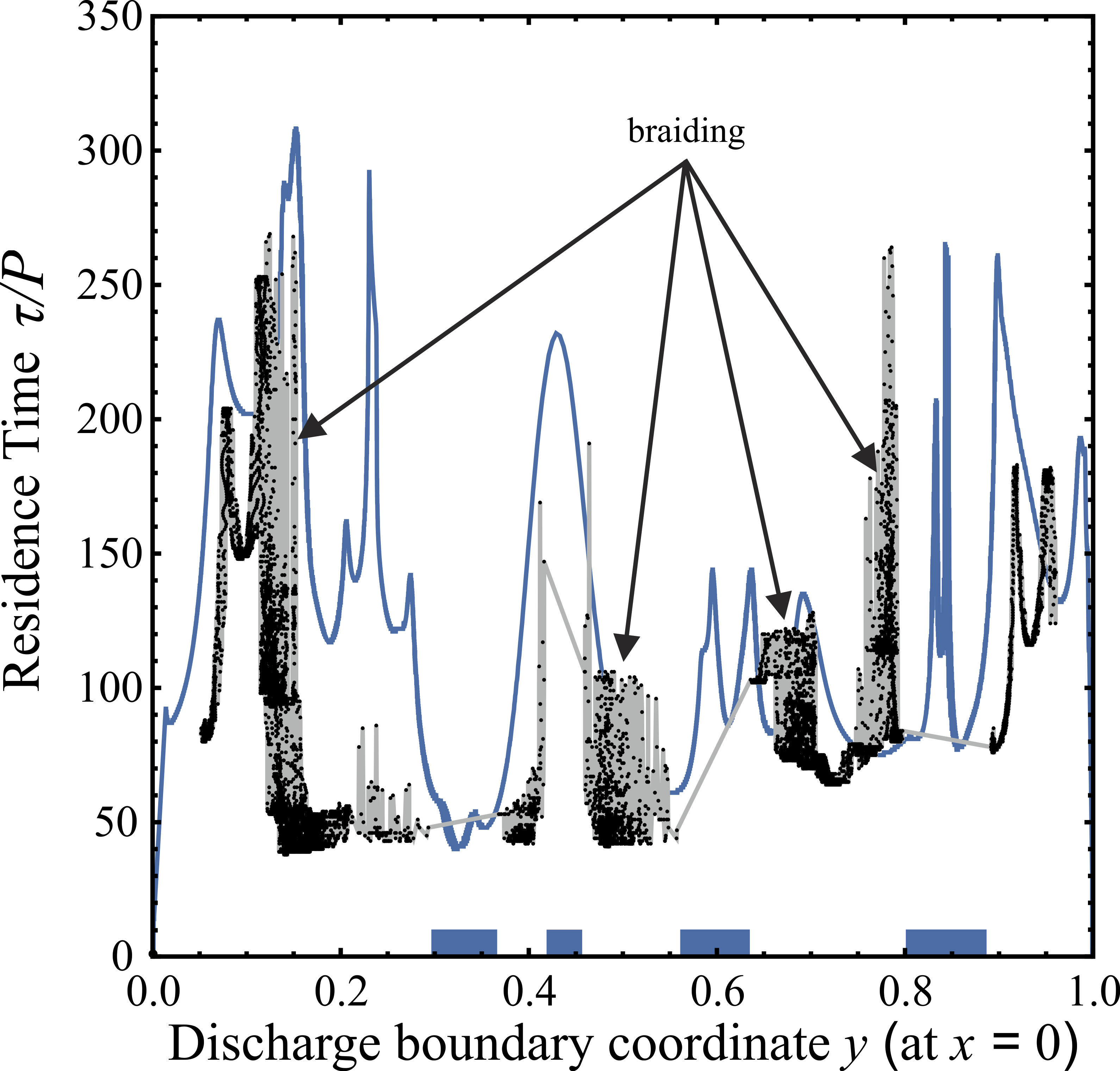}\\
  (a)\\[2em]
\includegraphics[width=0.98\columnwidth]{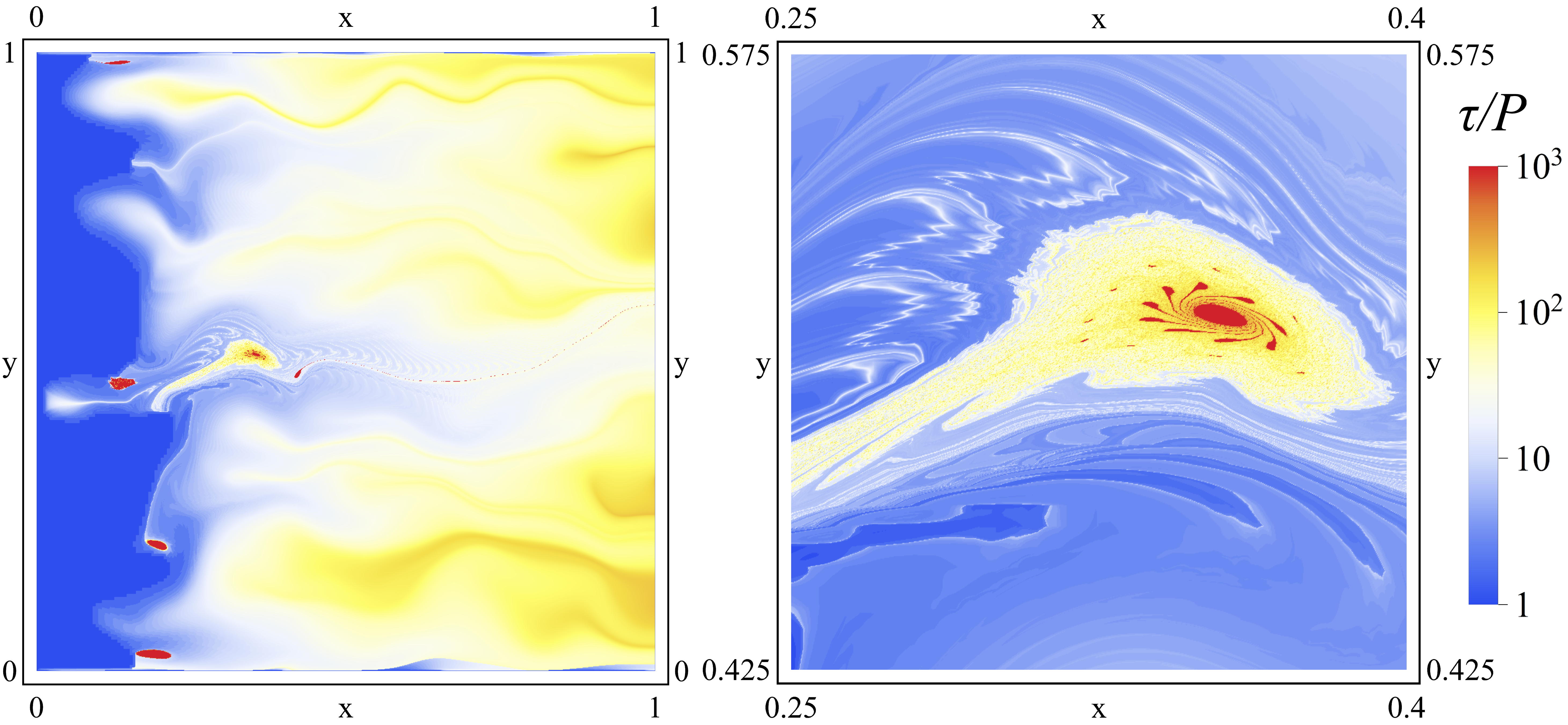}\\
(b)
\end{tabular}
\end{centering}
\caption{(a) The blue curve shows the RTD for the steady regional
  discharge flow, and the black dots indicate residence times for the
  tidally forced flow.  Blue bars indicate zones excluded to regional
  discharge.  Several zones of braided tidal discharge are noted.  (b)
  Maps of RTD for the transient tidal flow plotted in terms of initial
  position for a $1000\times1000$ grid of particles seeded over (left)
  the entire tidal domain and (right) a square region covering the
  structure centered at $(x,y)\approx(0.35, 0.51)$.  Particle tracking
  is performed up to $t/P=10^3$, hence red points have residence time
  $\tau>10^3 P$.}
\label{fig:rtd}
\end{figure}

Transport characteristics of coastal aquifers are primarily quantified in terms of the residence time distribution (RTD) associated with transport of fluid particles through the aquifer. Figure~\ref{fig:rtd}(a) shows the RTD ($\tau$) for $10^4$ fluid particles seeded in a flux-weighted distribution along the inland boundary ($x=1$) for a steady discharge flow (blue curve), $\mathcal{G}=0$, and for the transient tidal flow (black dots and grey lines), $\mathcal{G}>0$. Figure \ref{fig:rtd}(b) shows a map of the RTD for the transient tidal flow plotted as a function of initial position for a $1000\times 1000$ grid of fluid particles seeded across (center) the aquifer domain $\mathcal{D}$ and (right) a square region covering the largest mixing region of the flow.

The steady regional flow RTD shown in Figure \ref{fig:rtd}(a, blue
curve) is continuous and consists of well-defined peaks and troughs
that are controlled solely by the heterogeneity of the conductivity
field.  Conversely, the tidal RTD displays some regions with
well-defined, smooth, continuous peaks and troughs, but also other
zones of apparently stochastic nature where residence times vary
abruptly.  Highly resolved plots (not shown) of these
residence times indicate they are indeed smooth, but at very small
scales.  These stochastic zones are interpreted as intervals where
neighbouring flow paths are braided (entwined), as demonstrated in
Figure \ref{fig:flowpaths}(a), so that adjacent discharges at the
tidal boundary may originate from widely separated flow paths with
very different travel times.

The tidal RTD appears to be discontinuous at the macroscale in that there are large gaps in the discharge boundary position shown in in Figure~\ref{fig:rtd}(a).  These are indicated by straight grey lines between black points in the RTD trace and the blue bars along the horizontal axis. The RTD gaps correspond to "exclusion zones" along the tidal boundary ($x=0$), through which inland flow originating at the inland boundary never exits. Rather, the inland flow discharges in the gaps between these exclusion zones.  This unexpected result means that the combination of tidal forcing with aquifer heterogeneity can lead to significant intervals of the tidal boundary being inaccessible to the discharge of regional flow.  Below we offer a tentative explanation for the existence of exclusion zones.

Note that the RTD for particles seeded along the inland boundary ($x=1$) for the transient tidal flow (or steady regional flow) do not exhibit diverging residence times (even when much higher resolution of the RTD distribution is computed than that shown in Figure~\ref{fig:rtd}(a)), despite observations of such for flow reversal described in Subsection~\ref{subsec:flow_reversal}.  This is due to the fact that particles which are seeded at the inland boundary cannot enter these trapping regions for reasons which are explained in Subsection~\ref{subsec:Poincare}.  Conversely, Figure~\ref{fig:rtd}(b) shows that when particles are seeded over the entire flow domain, there are several distinct regions with very long RTDs but which are impervious to particles seeded at the inland boundary.  Detail of the largest long RTD ``island'' indicates a complex, fractal-like structure; the nature and origin of this structure is explained in Section~\ref{subsec:Poincare}.  The blue region ($\tau\leqslant P$) indicates the set of trajectories that exit the domain within one flow period; we denote this region the \emph{tidal emptying region}, and the rightmost boundary of this region as the \emph{tidal emptying boundary}.

\subsection{Poincar\'e sections}
\label{subsec:Poincare}
The RTD distributions in the previous subsection indicate the presence of complex transport and mixing dynamics within tidally forced aquifers. As indicated by Figure~\ref{fig:flowpaths}, plotting these complex flow trajectories results in a tangle from which it is difficult to discern any coherent structures or Lagrangian topology. A useful tool to aid such visualisation for periodic flows is the \emph{Poincar\'e section}, which allows direct visualisation of the Lagrangian topology of the flow field and interpretation of the Lagrangian kinematics within each topologically distinct region. A Poincar\'e section is formed by tracking a number of fluid particles via the advection equation (\ref{eq:pathline}), and recording all particle positions at every time period $\mathcal{P}$ of the flow.  The computation does not re-inject particles; once a particle exits the domain it is removed from the simulation. This \emph{stroboscopic map} essentially ``filters out'' all of the rapid particle motions between tidal forcing periods, leaving only the slow mean particle motion over each forcing period. Henceforth we shall refer to the flow averaged over a forcing period as the \emph{slow flow} of the aquifer. It can be shown that fluid transport in periodic flows can almost be completely understood solely in terms of this slow flow, hence Poincar\'{e} sections unveil the hitherto hidden transport structure of the aquifer. This approach resolves the Lagrangian topology of the flow into ``regular'' (i.e. non-chaotic) regions of the flow (with smooth, ordered particle paths) and chaotic mixing regions (with seemingly random particle locations).

\begin{figure}
\centering
\subcaptionbox{Steady flow solution}{\includegraphics[width=0.475\textwidth]{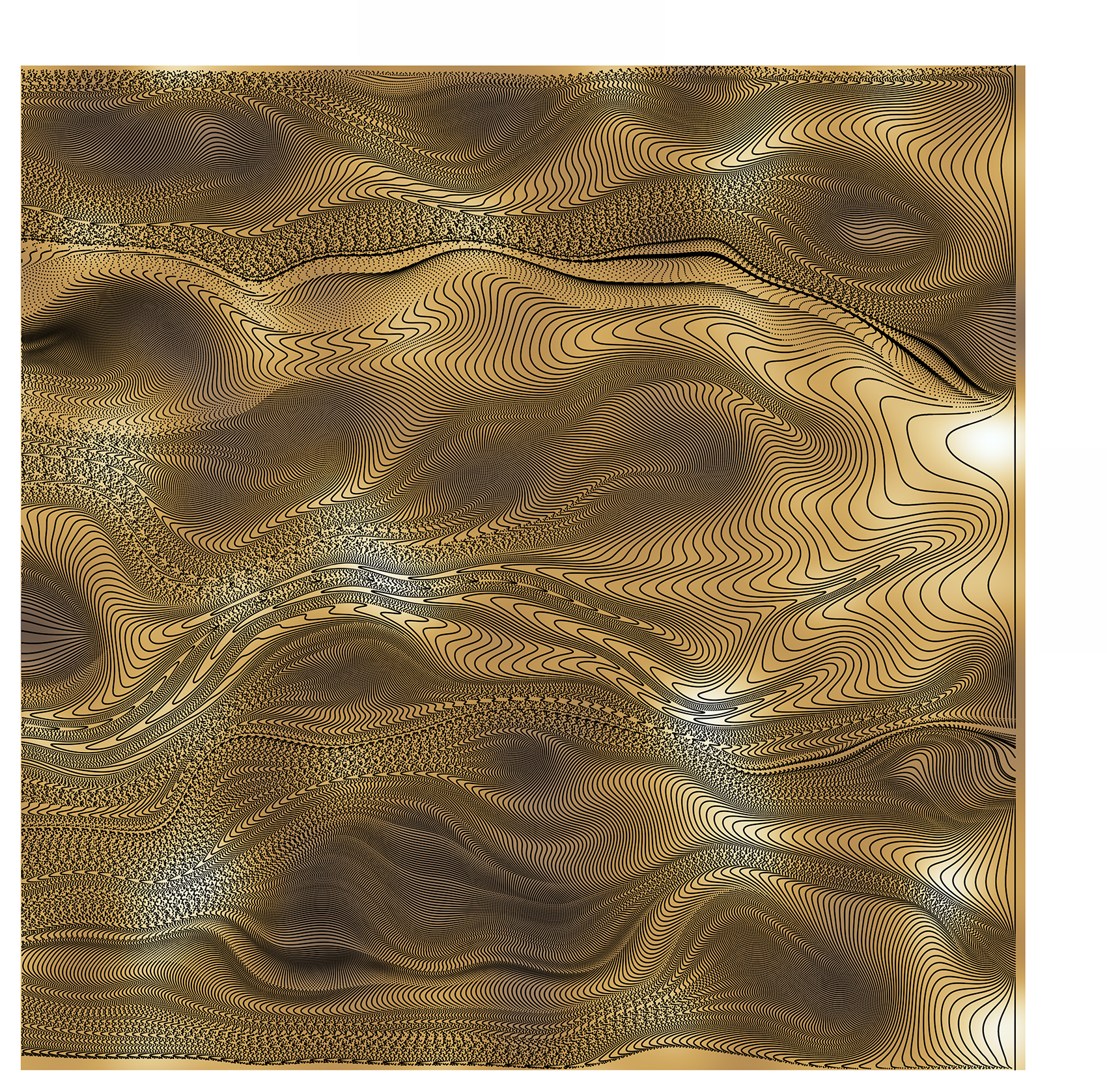}}%
\hfill
\subcaptionbox{Full tidal solution}{\includegraphics[width=0.475\textwidth]{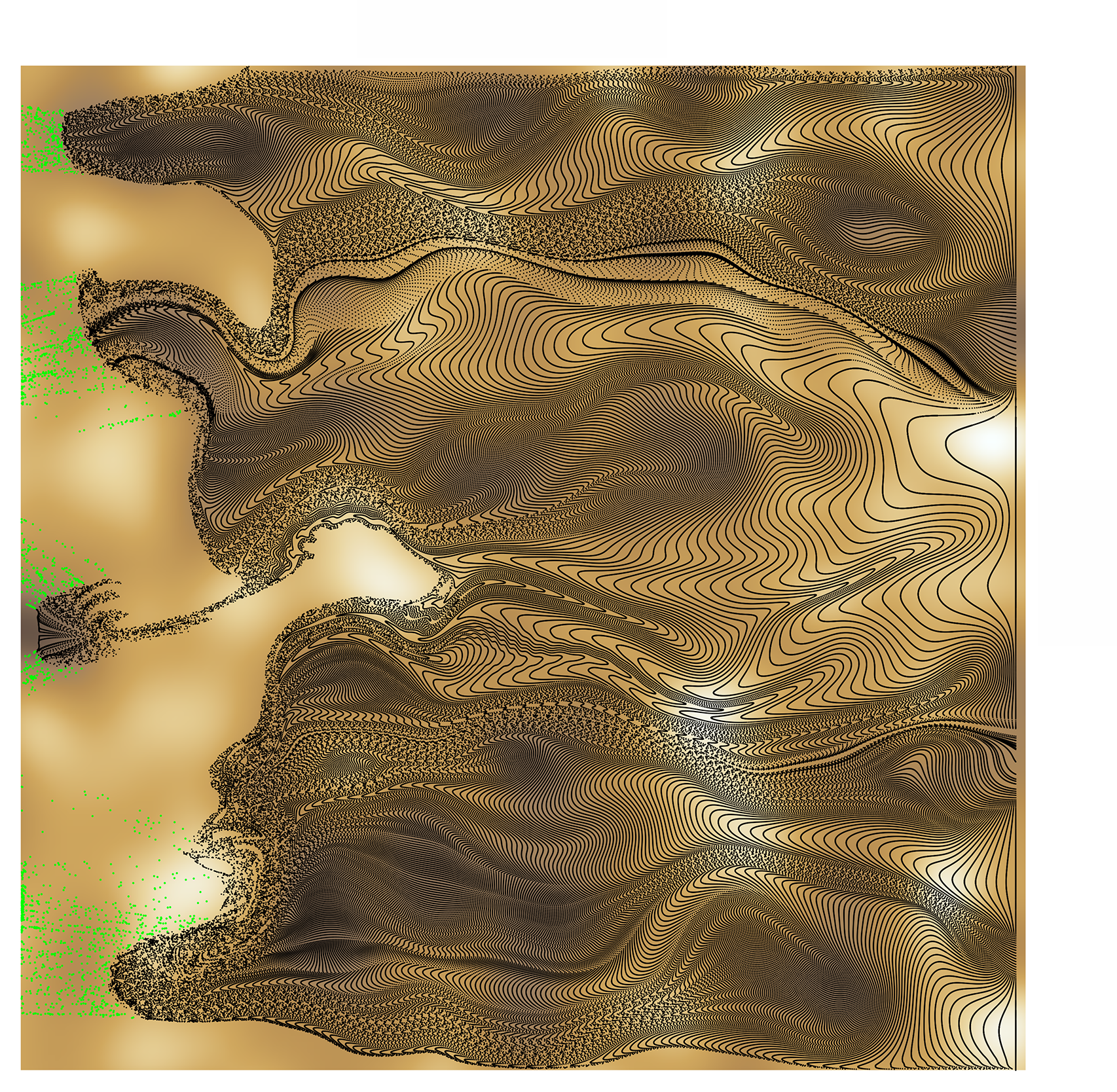}}%
\caption{Poincar\'e sections of regional discharge flux calculated with $\mathcal{P}$=\textit{P} for steady and for tidal flow cases. The tidal section also shows discharging points in green, calculated with $\mathcal{P}$=\textit{P}/10 from the particle locations at the last completed tidal period before discharge.}
\label{fig:psectionsx2}
\end{figure}

Figure~\ref{fig:psectionsx2} shows two Poincar\'e sections generated by releasing over 3,000 particles along a vertical line near the inland boundary $(x=1)$ for (a) the steady regional flow and (b) the transient tidal flow.  The structure of the Poincar\'e section for the steady flow arises solely from heterogeneity of the aquifer, where focusing of particles into (away from) high (low) permeability regions is apparent~\citep[see, e.g.][]{Trefry:2003aa}, along with differences in advective speed through these high/low permeability regions.  Whilst the Poincar\'e section of the tidal flow is almost identical to that of the steady flow in the vicinity of the inland boundary, significant deviations occur near the tidal boundary.  Most apparent is a large region (the \emph{tidal emptying region}, where $\tau<P$) near the tidal boundary which is devoid of fluid tracer particles seeded from the inland boundary.  Within this tidal emptying region there may be subregions where tracer particles may be (i) discharged in times less than a flow period (a \emph{discharge region}), (ii) enter from outside the aquifer domain ($x<0$) (an \emph{entry region}), or (iii) are trapped within a subregion indefinitely (a \emph{trapped region}).

\begin{figure}[t]
\centering
\includegraphics[width=0.98\columnwidth]{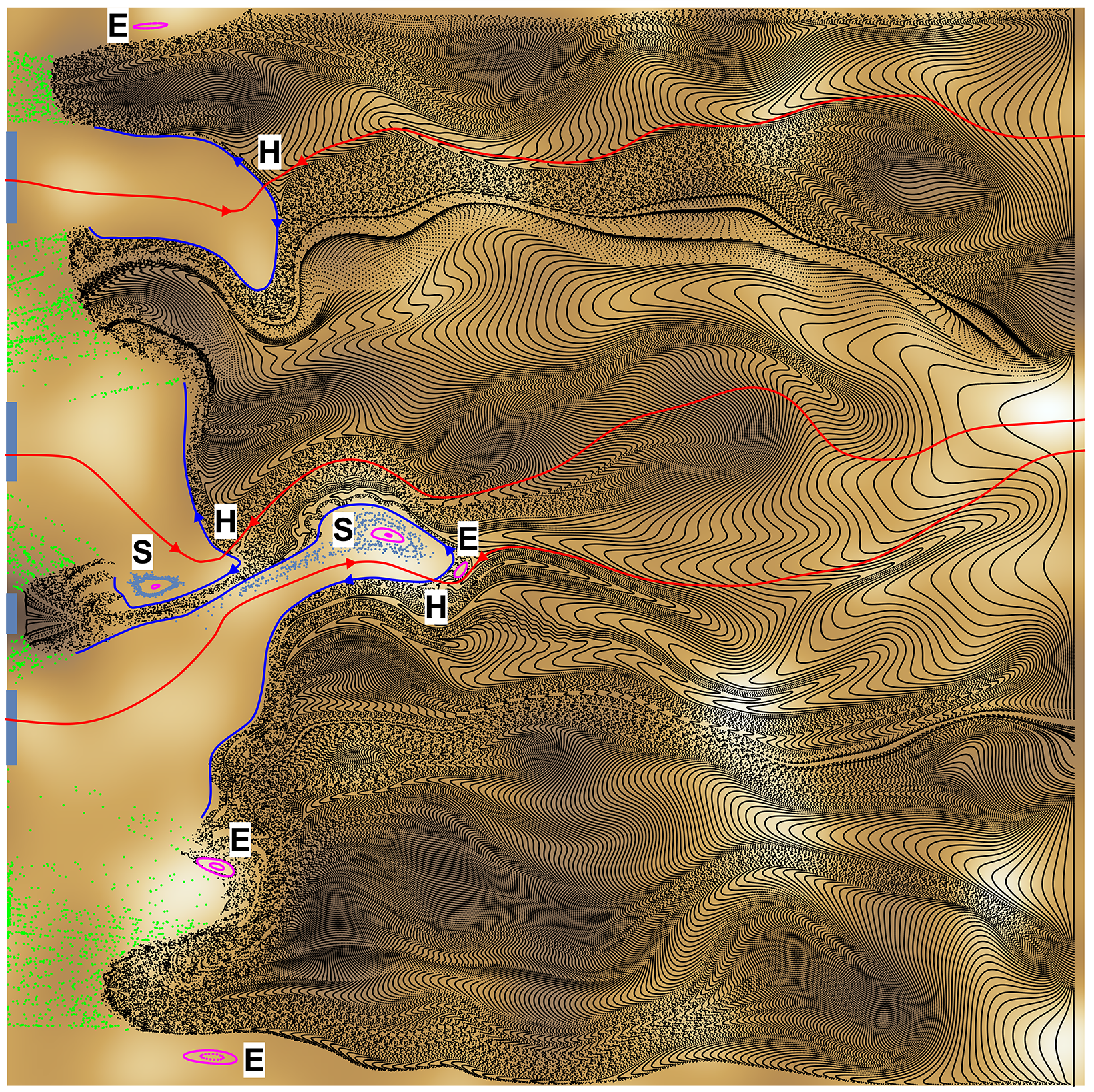}
\caption{Poincar\'e section of tidally forced aquifer showing
  hyperbolic (H) and elliptic (E) points, stochastic layers (S),
  stable (red lines) and unstable (blue lines) manifolds, KAM islands
  (magenta orbits), discharge points (green points) and exclusion
  zones (blue bars).}
\label{fig:annotatePoincare}
\end{figure}

Discharge regions are clearly illustrated by the green points in
Figure~\ref{fig:psectionsx2}(b), which indicate the locations of
particles that originated from the inland boundary ($x=1$) within the
tidal region at temporal increments of $t=P/10$. As shown, discharge
along the tidal boundary ($x=0$) coincides with gaps in the exclusion
zones shown in the RTD plot in Figure~\ref{fig:rtd}(a). This indicates
that the exclusion zones along the tidal boundary are associated with
the inflow of particles from the oceanic side of the tidal boundary
($x<0$, not modeled).  Although the dynamics in these discharge
regions appear to be rich (as indicated by the green points in
Figure~\ref{fig:annotatePoincare}), we do not consider them further in
this study.  Exclusion zones correspond to the blue bars and gaps in
the outflow RTD shown in Figure~\ref{fig:rtd}(a), and which themselves
correspond to inflow regions into the tidal emptying region.

When compared with the residence time distribution plot in
Figure~\ref{fig:rtd}(b), the tidal region appears to be mainly (but
not completely) comprised of particle initial positions with residence
times that are either less than one period ($\tau<P$) or have
diverging residence time ($\tau\rightarrow\infty$). The short
residence time regions indicate discharge or entry regions, whilst the
diverging residence times indicate trapped regions within the tidal
flow system.

\subsection{Elucidation of Lagrangian kinematics and Lagrangian topology}

Whilst these observations give a qualitative picture of the complex
transport and mixing dynamics in the tidal region, many features are
hidden due to selectivity from seeding particles only at
the inland boundary $(x=1)$.  To resolve the Lagrangian
kinematics of the flow fully, in Figure~\ref{fig:annotatePoincare} we
perform a closer inspection of the Poincar\'e section by seeding fluid particles
throughout the tidal region and resolving Lagrangian coherent structures such as periodic points and invariant manifolds.  Here we see, to our knowledge for the first time, how fluid transport due to tidal flows in heterogeneous aquifers is organized and the associated Lagrangian transport structure.  These are described in detail as follows.

\subsubsection{Periodic points}
To clearly elucidate the mixing and transport properties of the
aquifer, it is useful to find and classify the structures which
control the Lagrangian topology of the flow.  To understand these
structures it is necessary to introduce key concepts and nomenclature
from dynamical systems theory.  For periodic flows, key points that
organise the Lagrangian kinematics are the \emph{periodic
  points}~\citep{Ottino:1989aa} of the flow, i.e.\/ points that return
to the same position after one or more periods of the flow.  Points
that return after $p$ flow periods are denoted as period-$p$ points,
and typically lower-order (e.g. period-1, period-2) points play a
greater role in organising transport and mixing.  Such periodic points
are inadmissible in steady Darcy flow (see
Figure~\ref{fig:psectionsx2}a), and the presence of periodic points in
the tidally forced flow (see Figure~\ref{fig:annotatePoincare}) is
accompanied by a change in the flow topology from open, locally
parallel streamlines~\citep{Bear:1972aa} everywhere to the admission
of closed particle orbits and separatrices.  Such topological changes
bring about vast changes to transport.

Periodic points may be classified in terms of the net local fluid deformation that occurs after $p$-periods; it can be shown in two dimensions that this consists only of either local fluid rotation or local fluid stretching. Periodic points associated with such deformation are termed respectively \emph{elliptic} (E) and \emph{hyperbolic} (H) points; the former are associated with non-mixing, regular transport, whereas the latter may be associated with chaotic mixing. As shown in Figure~\ref{fig:annotatePoincare}, elliptic points are associated with non-mixing ``islands'' (formally known as KAM islands).

The exclusion zones penetrate far into the aquifer, each terminating at a \emph{hyperbolic point} (H) at the four-way intersection of associated stable (two) and unstable (two) manifolds. Unstable manifolds (departing blue curves) define the exclusion zone boundaries, while stable manifolds (arriving red curves) act as separatrices for slow fluid trajectories. It is important to remember that whilst structures in the Poincar\'{e} section govern the \emph{slow flow} of particles, complete trajectories of fluid particles include the fast oscillatory motion (as shown in Figure~\ref{fig:flowpaths}) between periods. 

The inland stable manifolds divide the regional discharge flow, deflecting to either side of the hyperbolic point. The stable manifolds that intersect the tidal boundary likewise divide the slow flow circulating within the exclusion zone into clockwise or anticlockwise slow motions. The nature of the exclusion zones is now apparent -- at high tides water enters the aquifer at the boundary and a fraction of this water performs long excursions into the aquifer, over many tidal periods, moving slowly towards the hyperbolic point before eventually returning to the boundary guided by the nearest unstable manifold. Three separate hyperbolic points are identified in the section, each with their own stable and unstable manifolds. In this way the aquifer contains significant volumes of fluid sourced from the tidal boundary and from the inland boundary; fluids within these two volumes are segregated by the unstable manifolds and do not intermingle until encountering the discharge region (indicated by green points). Note that stable/unstable manifolds cannot terminate in the fluid interior; the unstable manifolds which appear to terminate at the discharge regions in Figure~\ref{fig:annotatePoincare} do so as they have not been fully resolved.

\subsubsection{Elliptic points}
A number of \emph{elliptic points} (E, enclosed by magenta ellipses) are also identified in the Poincar\'{e} section. These occur both within the regional discharge flow and within the exclusion zones. The important dynamical characteristic of elliptic points is that they are surrounded by closed orbits of the slow motion, i.e. the orbits are fixed structures within which fluid circulates perpetually, with infinite residence time. These effects are not reflected in the residence time distributions initiated from the regional flow boundary (Figure~\ref{fig:rtd}a) since fluid in the elliptic orbits is never released to the discharge boundary (Figure~\ref{fig:rtd}b). The presence of elliptic points also indicates the potential for strong flow segregation in the natural tidal system. Figure~\ref{fig:annotatePoincare} shows the presence of elliptic points within the aquifer domain (four labeled E and two associated with stochastic layers), indicating the presence of trapping regions (KAM islands) which hold diffusionless fluid particles perpetually. Even in the presence of diffusion and hydrodynamic dispersion, these finite-sized regions can have a significant impact on solute transport~\citep{Lester:2014aa}.

\subsubsection{Chaotic saddles, stochastic layers and cantori}
\label{chaotic_saddles}

\begin{figure}[tp]
\begin{centering}
\begin{tabular}{c c}
\includegraphics[width=0.95\columnwidth]{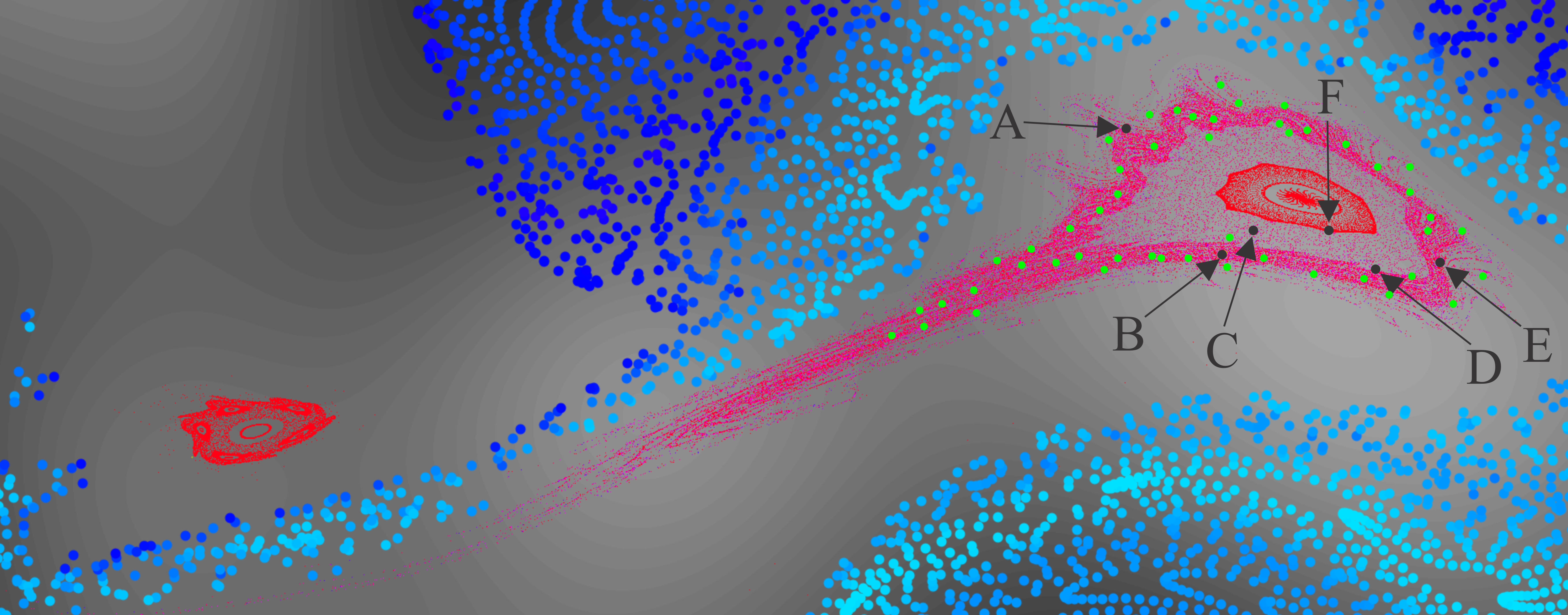}\\
(a)\\
\includegraphics[width=0.95\columnwidth]{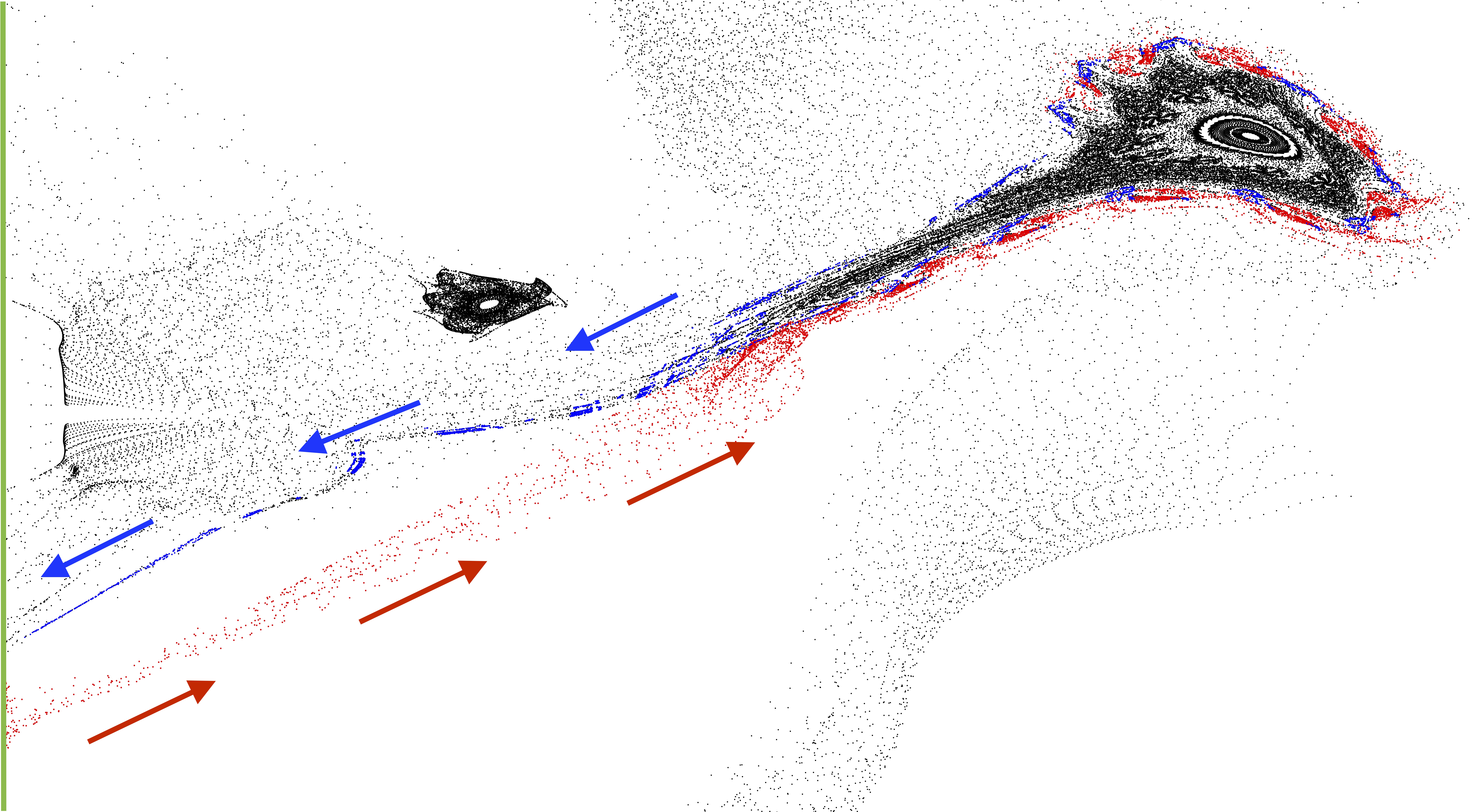}\\
(b)
\end{tabular}
\end{centering}
\caption{(a) Zoomed view of Poincar\'e section shown in Figure~\ref{fig:annotatePoincare}, illustrating the stochastic layers (pink dots) and elliptic orbits (red dots) in the mixing regions and regular orbits (blue dots) in the regular regions of the tidal domain. Points are coloured according to residence time, from the least (cyan) to the greatest (red). Roman letters, black and green dots indicate initial positions for FTLE calculations discussed in Subsection~\ref{subsec:ftle}. (b) Detail of Poincar\'e sections for the mixing regions and surrounding non-chaotic structures in (a) as (black dots), showing points associated with stable (red) and unstable (blue) manifolds, respectively, entering and leaving the aquifer domain via the tidal boundary (green line). The chaotic saddle is  the intersection of these manifolds, which surround the elliptic orbits (KAM islands), cantori and stochastic layers.}
\label{fig:zoomstochastic}
\end{figure}

Two \emph{stochastic layers} are also identified, labeled by S and indicated by blue points. Each stochastic layer surrounds a set of (magenta) elliptic orbits around an elliptic point (unlabeled). Even though the inner elliptic orbits are perfectly closed, the stochastic layers consist of assemblages of finite orbits where fluid is trapped to circulate the elliptic point for long (e.g. hundreds or thousands of tidal periods) but apparently random circulation times, before being released back into the influence of the nearby unstable manifold for eventual discharge to the tidal boundary.  Formally, stochastic layers arise from \emph{cantori} which are fractal distributions of
elliptic points. Figure \ref{fig:zoomstochastic} shows a high resolution Poincar\'e section focused on the two stochastic layers in Figure \ref{fig:annotatePoincare}. This high resolution section provides a much clearer (though still imperfect) picture of the rich dynamical nature of the stochastic layer and associated cantori.

All of the stable and unstable manifolds shown in Figure~\ref{fig:annotatePoincare} that organise the basic Lagrangian topology form smooth connections and so do not generate chaotic mixing.  However, there also exist transverse intersections between stable and unstable manifolds (such as those shown in Figure~\ref{fig:VonKarman}(c)) that give rise to the chaotic saddles and mixing regions discussed in Section~\ref{subsec:flow_reversal}.  Figure~\ref{fig:zoomstochastic} show high resolution Poincar\'e sections focused on the two largest mixing regions shown in Figure \ref{fig:annotatePoincare}. 

Figure~\ref{fig:zoomstochastic}(a) shows the stochastic layer (pink points) that surrounds the series of nested elliptic orbits (KAM islands) (red points), and the non-chaotic (regular) regions are denoted by cyan and blue points.  These points are coloured with respect to the logarithm of residence time, where cyan/blue points leave the domain after $\mathcal{O}(10)$ periods, pink points $\mathcal{O}(10^2)$ periods and red points stay within the domain indefinitely.  These transport kinematics correspond to the residence time distributions shown in Figure~\ref{fig:rtd}(c), where indefinitely-trapped particles in the KAM islands and cantori are surrounded by long-lived (but finite-time) orbits in the stochastic layer.

Figure~\ref{fig:zoomstochastic}(b) illustrates the stable (red) and
unstable (blue) manifolds (shown as discrete points rather than the
continuous lines shown in Figure~\ref{fig:VonKarman}(c)).  For the
aquifer model under consideration, the stable and unstable manifolds
respectively enter and leave the aquifer domain via the tidal boundary
($x=0$) rather than having the stable manifold enter from upstream as
shown in Figure~\ref{fig:VonKarman}(c).  It is presently unknown
whether this behaviour is universal to all chaotic saddles in tidally
forced aquifers.  The chaotic saddle is formed by the transverse
intersection of the stable and unstable manifolds which both enter
from the tidal boundary $(x=0)$: the stochastic layer is formed
by the chaotic mixing dynamics associated with the heteroclinic tangle
between the stable and unstable manifolds, which leads to a fractal
(spatial) distribution of residence times in the stochastic layer, as
described in Section~\ref{chaotic_saddles}.

The interplay of transient forcing, compressibility and heterogeneity
in coastal aquifers leads to complex transport dynamics and a rich
Lagrangian topology.  Partitioning the Lagrangian topology into
distinct regions (which include particle trapping, mixing, particle
inflow and outflow) and analysis of the transport within these regions
gives an overview of the complex Lagrangian kinematics within tidally
forced heterogeneous aquifers.  In Section \ref{sec:field}, we discuss
the practical implications of these complex transport dynamics upon
solute mixing, transport and chemical reactions.

\subsection{Finite-time Lyapunov exponents}
\label{subsec:ftle}

Chaotic advection 
is characterised by the exponential stretching of fluid material elements, 
where the stretching rate is characterised by the (infinite-time) Lyapunov exponent, defined as
\begin{equation} \label{eqn:inf_time_lyapunov}
\Lambda_\infty\equiv\lim_{t\rightarrow\infty}\frac{1}{t}\ln\frac{\delta l(t)}{\delta l(0)}, 
\end{equation}
where $\delta l(t)$ is the length of an infinitesimal fluid line element advected by the flow. A positive Lyapunov exponent $\Lambda_\infty>0$ indicates the presence of chaotic advection, and the magnitude of $\Lambda_\infty$ indicates the rate of exponential stretching. For closed flows (whether steady or unsteady), the Lagrangian flow domain can be divided into topologically distinct regions which are either regular (non-chaotic, $\Lambda_\infty=0$) or chaotic ($\Lambda_\infty>0$).

As fluid elements can flow into and out of mixing regions in open flows, it is more useful to characterise deformation of fluid particles in terms of the finite-time Lyapunov exponent (FTLE, $\Lambda$) which quantifies the maximum (i.e. maximum over all possible initial orientations) deformation of an infinitesimal fluid line element (see Appendix~\ref{app:FTLE} for details). In the limit of infinite residence time the FTLE of an orbit converges to the infinite-time Lyapunov exponent of the chaotic saddle (which does not flow out of the mixing region), which is also equivalent to the ensemble average of the FTLE over many orbits over finite time.

\begin{figure}[t]
\centering
\includegraphics[width=0.65\columnwidth]{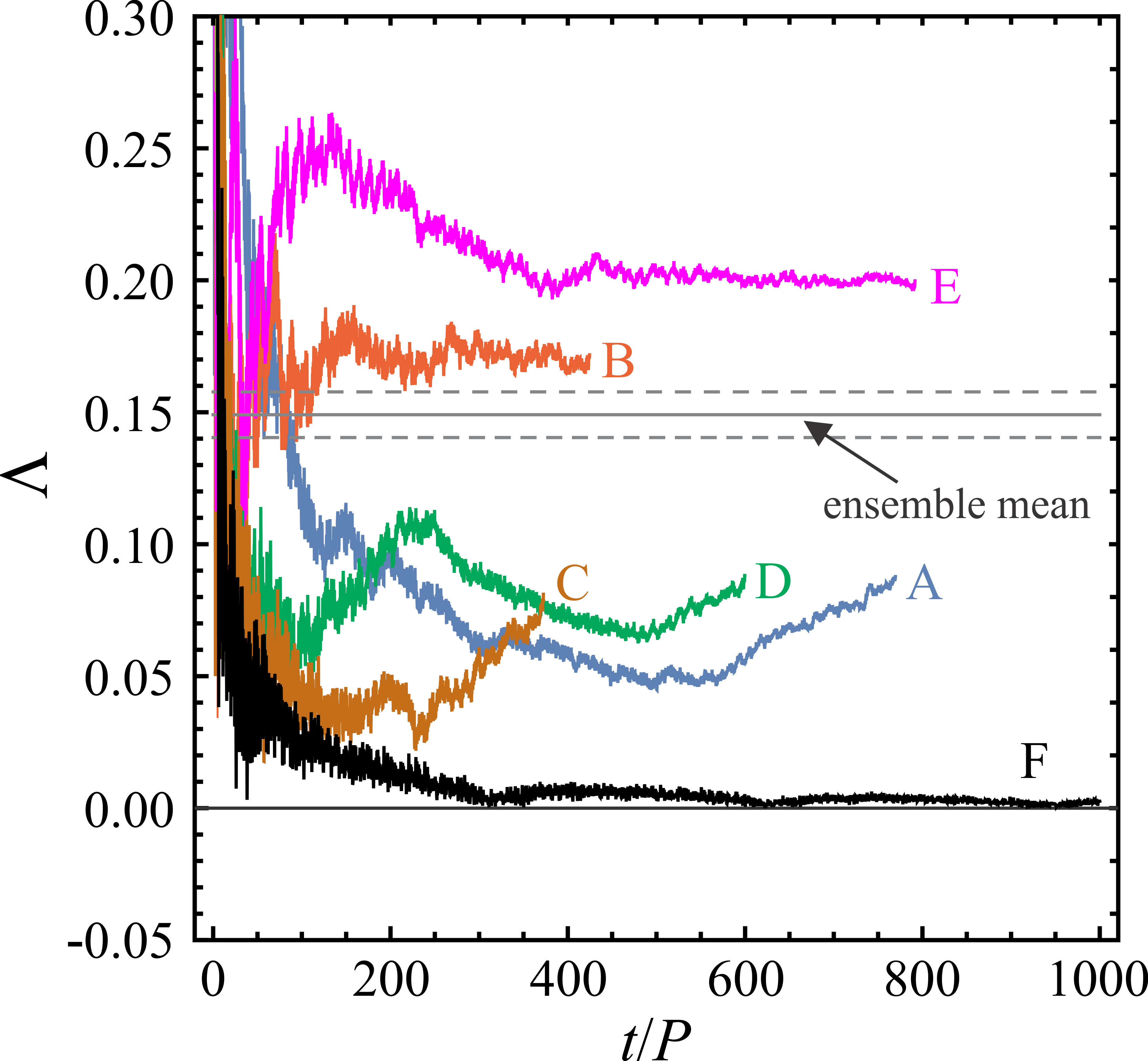}
\caption{Finite-time Lyapunov exponents calculated from five
  arbitrarily chosen starting locations in the stochastic layer (see
  Figure \ref{fig:zoomstochastic}).  All traces eventually exit at the
  tidal boundary, except trace F (KAM orbit) which does not terminate
  and was truncated for display.  The ensemble mean and standard error
  bounds are indicated.}
\label{fig:FTLEtraces}
\end{figure}

Figure \ref{fig:FTLEtraces} provides FTLE traces calculated from arbitrarily chosen starting locations in the large stochastic layer of Figure \ref{fig:zoomstochastic}. The FTLE traces are noisy at short times but become smoother after several hundred $P$ elapsed time. The stochastic layers circulate fluids entering the domain at the tidal boundary before releasing them for subsequent discharge. The recirculation times form a random process depending on particle starting location. Particles at locations A--E all eventually discharge, whereas particles at F circulate indefinitely ($> 10^{3} P$). The finite natures of trajectories A--E cause short-term shifts of the FTLE traces as the escaping particles encounter a finite sequence of different Lagrangian regions on their way to the boundary. Thus the FTLE traces in Figure \ref{fig:FTLEtraces} are not monotonically convergent to limiting values, as is common for evaluations of infinite-time Lyapunov exponents ($\Lambda_{\infty}$). To overcome this estimation problem we calculated FTLE values for a random ensemble of 42 starting locations (green and black dots in Figure \ref{fig:FTLEtraces}), gaining an ensemble mean FTLE  $\approx 0.149$ with low standard error. For comparison, \citet{Lester:2013ab} show that pore-scale branching networks lead to an infinite-time Lyapunov exponent of $\Lambda_{\infty} \approx 0.12$. Thus our ensemble of (finite) open flow trajectories in the stochastic layer displays a positive mean FTLE indicative of chaos, while the infinite KAM orbit (F) displays only algebraic $(\Lambda=0)$ deformation.

\section{Discussion and physical relevance}
Results from the previous sections have clearly established the presence of complex transport phenomena and rich Lagrangian topology in tidally forced aquifers. In this section we consider the impacts of these transport dynamics on transport, mixing and reactions, and we place the example problem in the context of a selection of relevant field studies.

\subsection{Implications for transport and reaction}
\label{subsec:implications}
Groundwater discharge is a prime vector for contaminant migration and environmental impact. However it is beyond the scope of this paper to extend our Lagrangian analysis to dispersive/diffusive systems, although this is clearly an important direction for future research. Here we limit our comments to observations of Lagrangian phenomena in our simulations that impinge upon the utility of conventional groundwater quality characterization activities.

\subsubsection{Residence time distributions - segregation and singularity}

In our example problem we saw that the residence time distribution of fluid in the aquifer domain was singular (infinite) at several locations controlled by periodic points. Furthermore, regional flux was diverted into several discrete discharge zones along the tidal boundary, and the fluids discharging in these zones had undergone complicated braiding dynamics which may mix and mask geochemical signatures. At other locations along the boundary there were intervals of recirculation (with widths of the order of the log-conductivity correlation length $\lambda$) where tidal fluids embarked on long incursions into the aquifer domain, lasting many tidal periods, before eventually returning and discharging at the boundary. Thus, taken as a whole, fluids discharging along the tidal boundary may have vastly different (and possibly indeterminate) geochemical and chronological origins. This behaviour is fundamentally different to the familiar steady discharge dynamics in heterogeneous domains (see the blue curve in Figure \ref{fig:rtd}). Conventional groundwater sampling techniques near discharge boundaries, e.g. multilevel samplers and spear probes, may variously source fluids from an ensemble of histories and qualities. High sampling density, in space and time, may be the best practical defense against these unpredictable effects in poorly characterized aquifers.

\subsubsection{Flow reversal and traceability}

The origin of these complicated transport dynamics can be traced back to reversal of the Darcy flux vector during the tidal forcing period over a significant proportion of the aquifer domain (see Figure~\ref{fig:canonicalzone}), leading to a fundamental change in the Lagrangian topology from open, parallel streamlines to the formation of periodic points, separatices, non-mixing islands and chaotic saddles (see Figures~\ref{fig:psectionsx2}, \ref{fig:annotatePoincare}). Moreover, our observations of punctuated inflow/outflow regions along the tidal boundary ($x=0$) significantly alter understanding of transport in these regions, especially from the perspective of, for example, salt water intrusion. Such complex transport dynamics are incompatible with many solute transport and biogeochemical reaction modeling studies which tend to rely on the identification of simple mean flow paths along which complicated mass balance and reaction kinetic calculations can be made. Conceptually, our results suggest that determination of the fate (provenance) of reacting fluids simply by (back-) tracking along an approximate steady streamline may well be a source of significant model error in coastal discharge environments. That is, determination of mean fluid flow paths for geochemical reaction modeling and transport may not be appropriate or even possible in some coastal settings. In addition, it has been shown~\citep{Tel2005} that the complex mixing dynamics that arise within chaotic regions can profoundly alter a wide range of chemical and biological reaction rates.

The highly oscillatory nature of particle motion near the tidal
boundary (see Figure~\ref{fig:flowpaths}) can also significantly skew
estimates of tracer migration speed. Although the net displacement of
a particle over a tidal period $P$ may be small, the mid-cycle
particle displacement can be much larger.  Noting that most tidal
spectra are multi-modal and that the Townley number $\mathcal{T}$ is
frequency-dependent, the appropriate sampling frequency may be a
function of sampling location in the aquifer. It is clear from our
results that tracer test analysis in strongly tidally influenced
aquifers is potentially fraught.

\subsection{Potential for chaos in field settings}
\label{sec:field}

\begin{figure}
\centering
\includegraphics[scale=0.3]{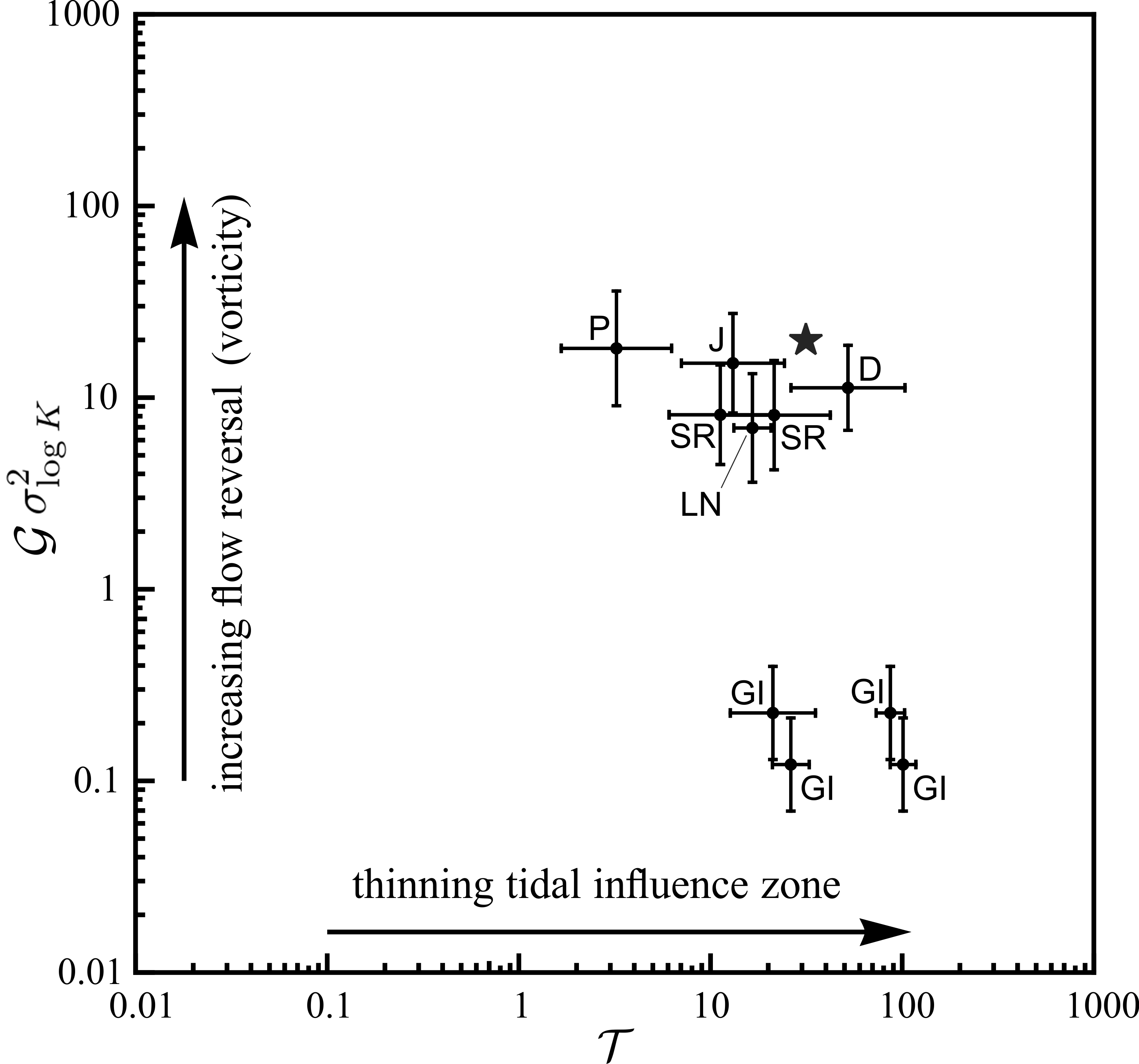}
\caption{Location of the present example system (star) in $\mathcal{T}$ -- $\mathcal{G}\,\sigma^{2}_{\textrm{log}K}$ parameter space. Locations of some field studies reported in the literature are provided for reference (see text for details).}
\label{fig:TRspace}
\end{figure}

It is important to consider how the dimensionless parameter set
$\mathcal{Q}$=($\mathcal{T}$, $\mathcal{G}$,
$\mathcal{C}$)=(10$\pi$,10,0.5) and conductivity parameters
($\sigma^2_{\log K}=2$, $\lambda=0.05$) of our example problem relates
to those of typical coastal aquifers.  Noting that the product
$\mathcal{G}\,\sigma^2_{\log K}$ correlates well with the density of
canonical flux ellipses in the tidally active zone, in order to assess
the relevance of these parameter values to field studies we estimate
$\mathcal{T}$ and $\mathcal{G}\,\sigma^2_{\log K}$ values from a set
of published coastal aquifer studies. The study locations are the
Dridrate Aquifer, Morocco (D, sand and limestone)
\citep{Fakir:2003aa}, Garden Island (GI, sand and limestone)
\citep{Trefry:2004aa}, Jervoise Bay (J, sand and limestone)
\citep{Smith:2001aa} and Swan River (SR, sand and clay)
\citep{Smith:1999aa}, all in Western Australia, Largs North, South
Australia (LN, sand and clay) \citep{Trefry:1998aa}, and Pico Island,
Azores (P, basalts) \citep{Cruz:2001aa}. Figure~\ref{fig:TRspace}
plots these studies in $\mathcal{T}$--$\mathcal{G}\,\sigma^2_{\log K}$
space with error bars indicating uncertainty in hydrogeological
parameter estimates. The GI, SR and J studies include multiple data
points due to the use of frequency-resolved techniques. Estimates of
$\sigma_{\textrm{log} \kappa}^{2}$ are rare in tidal analyses although
suitable estimation techniques exist \citep[see
e.g.][]{Trefry:2011aa}. The following ranges of
$\sigma_{\textrm{log} \kappa}^{2}$ were assumed: 0.5--2.5 (sand and
limestone, sand and clay), 2.0--6.0 (basalts).
Figure~\ref{fig:TRspace} shows that the example problem (star) lies
well within the range of $\mathcal{T}$ values of the field studies but
at the upper end of the estimated flow reversal
$(\mathcal{G}\,\sigma^2_{\log K})$ values.

The tidal compression ratio $\mathcal{C} = S\!_{s} \, g_{\text{p}}/\varphi_{\text{ref}}$ is also a key parameter. As porosity $\varphi_{\textrm{ref}}$ is not always well characterized in tidal aquifer studies,  we resort to literature values for soil physical parameters \citep[][]{Domenico:1965aa,Morris:1967aa}. For sand and limestone soils, $\varphi_{\textrm{ref}}$ ranges from 0.2--0.5, while for sand and clay soils the range is 0.15--0.4. Basalts, often with significant fracture porosity, may show $\varphi_{\textrm{ref}}$ in the range 0.03--0.35. Specific storages $S_{s}$ typically range up to $10^{-2}\,\textrm{m}^{-1}$ (sand and limestone, sand and clay) and $10^{-4}\,\textrm{m}^{-1}$ (fractured rock), while tidal amplitudes can be as large as 10 m. With these ranges the maximum $\mathcal{C}$ values that can be anticipated are approximately 0.05 (sand and limestone), 0.4 (sand and clay) and 0.02 (basalts), but lower tidal amplitudes will reduce these proportionately. Thus we see that the example problem, with $\mathcal{C} = 0.5$, is at the high end of likely values in field settings, closest to a clay and sand matrix influenced by high tidal amplitude. It is beyond the scope of this paper to identify specific instances of Lagrangian chaos in coastal aquifers, but the preceding analysis shows that the parameter choices in the example problem are not far removed from reality and there is good prospect of coherent Lagrangian structures being present near discharge boundaries in many field settings.

However, whether such structures may be detected in field
investigations is another matter altogether. Submarine groundwater
discharge (SGD) mapping techniques are directly relevant to the
detection of Lagrangian structures in coastal systems, but the
uncertainties involved in SGD are significant. Firstly, coastal
discharge zones often display complex geomorphology, topography and
fully three-dimensional heterogeneity at a range of spatial
scales. This complexity is sufficient to render the SGD measurement
task onerous, even at relatively poor spatial resolution. Once
gathered, the measurements show significant spatial and temporal
variability due to geological structure, temporal variations in flow
regime, and measurement uncertainty. The measurement variability is
often rationalized in terms of (unresolved) preferential flow paths
\citep[see, for example][]{Hosono:2012aa,Smith:2003aa} although other
mechanisms including density effects, wave action \citep{Smith:2009aa}
and tidal and weather system influences are also reported to be
important \citep{Kobayashi:2017aa}. Our results indicate that tidal
forcing introduces a new dynamical mechanism for generating
preferential flow paths via the establishment of interior hyperbolic
points, resulting in segregation of regional flow into discrete
discharge zones.  It remains to be seen whether techniques can be
developed to isolate the presence and quantify the effects (on flow,
mixing and reaction) of Lagrangian structures in coastal
systems. Nevertheless, the potential for Lagrangian chaos must be
recognized and assessed.

\section{Conclusions}
In this paper, motivated by earlier studies in coastal groundwater hydraulics and biogeochemical transport, we have considered the dynamics and kinematics of (dispersion-free) flow in a simple 2D time-periodic groundwater discharge system, which represents a broad class of natural (unpumped) hydrogeological environments. Using standard linear poroelastic theory, spectral solutions to the linear Darcy flow equations were obtained using a finite-difference scheme coupled with a self-consistent streamfunction method which resulted in a highly accurate, time-periodic velocity field. Spatial heterogeneity of aquifer conductivity was accounted for explicitly, yielding complex and coupled spatio-temporal variations in head, porosity, flux and velocity.

We have, for the first time, uncovered the underlying transport structure of tidally forced aquifers. Rather than study the advection-diffusion of a passive scalar, we advect diffusionless tracer particles to uncover this non-trivial transport structure. Resolution of the Poincar\'{e} section of the example groundwater flow reveals a rich Lagrangian topology, with features that include:
\begin{enumerate}
	\item \emph{Elliptic islands} - isolated regions of the flow which trap fluid particles indefinitely
	\item \emph{Chaotic saddles} - isolated regions of the flow which involve rapid mixing and augmented reaction kinetics 
	\item \emph{Flow segregation} - division of the flow domain into distinct subdomains
	\item \emph{Tidal inflow/outflow} - division of the tidal boundary into distinct inflow or outflow regions
\end{enumerate}
This complex transport structure has profound ramifications for flow, transport and reaction in time-periodic groundwater discharge systems, and changes our understanding of transport and mixing processes (such as saltwater intrusion) in these systems at a fundamental level.  The interplay of this diffusionless transport structure and dispersive processes represents a pressing direction for future research.

We show that the interplay of tidal forcing and aquifer heterogeneity is responsible for these complex Lagrangian kinematics. Specifically it is the presence of flow reversal points in the aquifer that are directly responsible for the change in flow topology from everywhere parallel streamlines (associated with steady Darcy flow) to the presence of closed flow path lines, periodic points and separatrices. This change in flow topology fundamentally alters the transport characteristics of the aquifer, leading to indefinite trapping of fluid elements in circulation regions and segregation of the flow domain into distinct regions.

We also show that non-zero aquifer compressibility can lead to a
further topological bifurcation to form chaotic saddles: localised
regions of chaotic advection that involve rapid mixing and augmented
reaction kinetics.  These chaotic regions augment transport and mixing
dynamics of the aquifer.  Our example problem used a relatively high
compressibility value (compared to typical field values) and we are
presently scanning the lower compressibility values to determine the
parametric extent of this chaotic behaviour.

The results presented in this paper demonstrate that a conventional
Darcian groundwater flow system has the potential to admit complex and
possibly chaotic flows as long as two preconditions are met: a
heterogeneous aquifer domain and at least one time-periodic discharge
boundary.

\appendix

\section{Method for self-consistent, physically constrained flows} \label{app:numerical_ce}
When considering the dynamical aspects of flow fields, evaluations of velocity vectors for particle tracking purposes require greater accuracy than is commonly provided in many groundwater modeling investigations. This is because the identification of Lagrangian structures can require a combination of long particle travel times and fine spatial resolution. Numerical calculations of the Darcy flux vector at arbitrary locations in the problem domain can be problematic since calculations require both differentiation and interpolation of discretised variables ($\kappa$ and $h$). Here we describe a numerical method to solve for the tidal heads via a finite difference approach, and then generate heads, fluxes and porosities in continuous forms consistent with the relevant physical constraints (\ref{eq:cedeform}) and (\ref{eq:nvsh}). These continuous variables allow the fluid velocity field to be evaluated to an accuracy sufficient to support the required analyses of Lagrangian flow characteristics. A \emph{Mathematica} version 11 notebook containing the full numerical implementation is available from the authors on request.

The steady and periodic head solutions of (\ref{eq:Helmholtz_nondim}) and (\ref{eq:hmbc_nondim}) calculated according to the numerical scheme in \citet{Trefry:2009aa} are presented as head distributions $h_{s}$ and $h_{p}$ evaluated on a regular (square) finite difference grid with spatial increment $\Delta$. The corresponding flux distributions $\textbf{q}_{s}$ and $\textbf{q}_{p}$ are constructed by finite differences according to
\begin{eqnarray} \label{eq:qfd}
\textbf{q}_{s}(\textbf{x}_{m,n}) &=& \left( -\bm{\kappa}^{+}_{x}\, \left[ h_{s}(\textbf{x}_{m+1,n}) - h_{s}(\textbf{x}_{m,n})\right], -\bm{\kappa}^{+}_{y}\, \left[h_{s}(\textbf{x}_{m,n+1}) - h_{s}(\textbf{x}_{m,n})\right]\right)/\Delta \nonumber\\
\textbf{q}_{p}(\textbf{x}_{m,n},t) &=& \dfrac{e^{i \omega t}}{\Delta} \left( -\bm{\kappa}^{+}_{x}\, \left[ h_{p}(\textbf{x}_{m+1,n}) - h_{p}(\textbf{x}_{m,n})\right], -\bm{\kappa}^{+}_{y}\, \left[h_{p}(\textbf{x}_{m,n+1}) - h_{p}(\textbf{x}_{m,n})\right]\right) \nonumber\\
\end{eqnarray}
where $m$ and $n$ index the finite difference nodes in the $x$ and $y$ directions, respectively, and $\textbf{x}_{m,n} = (x_{m},y_{n})$. For the present 2D problem the mid-nodal conductivity values may be estimated either by harmonic averages:
\begin{eqnarray} \label{eq:Kharm}
\bm{\kappa}^{+}_{x} &=& 2\, \bm{\kappa}_{m+1,n}\, \bm{\kappa}_{m,n}/(\bm{\kappa}_{m+1,n}+\bm{\kappa}_{m,n}) \nonumber\\
\bm{\kappa}^{+}_{y} &=& 2\, \bm{\kappa}_{m,n+1}\, \bm{\kappa}_{m,n}/(\bm{\kappa}_{m,n+1}+\bm{\kappa}_{m,n}) 
\end{eqnarray}
or geometric averages:
\begin{eqnarray} \label{eq:Kgeom}
\bm{\kappa}^{+}_{x} &=& \sqrt{\bm{\kappa}_{m+1,n}\, \bm{\kappa}_{m,n}} \nonumber\\
\bm{\kappa}^{+}_{y} &=& \sqrt{\bm{\kappa}_{m,n+1}\, \bm{\kappa}_{m,n}}. 
\end{eqnarray}
In this work geometric averaging is used exclusively. The flux estimation scheme (\ref{eq:qfd}) provides error residuals identical to those of the head solution. The nodal porosity is expressed as
\begin{eqnarray} \label{eq:porosityfd}
\varphi(\textbf{x}_{m,n},t) &=&  \varphi_{\textrm{ref}} + S \left(h_{s}(\textbf{x}_{m,n}) - h_{\textrm{ref}} + h_{p}(\textbf{x}_{m,n}) e^{i \omega t}\right) \nonumber\\
&=& \varphi_{s}(\textbf{x}_{m,n}) + \varphi_{p}(\textbf{x}_{m,n},t)
\end{eqnarray}

Consider the continuity equation (\ref{eq:cedeform}). This must be satisfied at all locations (not just at the finite-difference nodes) and times in the model system in order for fluid mass to be conserved. Noting that, formally, the Darcy flux $\textbf{q}$ and porosity $\varphi$ satisfy (\ref{eq:periodicq}) and (\ref{eq:nvshperiodic}), respectively, we see that the continuity equation can be expressed as (involving continuous quantities):
\begin{equation}
\label{eqn:app_C1}
\nabla \cdot (\textbf{q}_{s} + \textbf{q}_{p} e^{i \omega t}) + \dfrac{\partial(\varphi_{s}+\varphi_{p} e^{i \omega t})}{\partial t} = 0
\end{equation}
Expanding this equation and noting the time-independence of the results yields the following two identities for exactly mass-conserving flow in our model periodic system:
\begin{eqnarray} \label{eqn:app_C2}
\nabla \cdot \textbf{q}_{s} = 0 \nonumber\\
\nabla \cdot \textbf{q}_{p} + i \omega \, \varphi_{p} = 0.
\end{eqnarray}
This result shows that the periodic component of the Darcy flux induces a bounded oscillation in the local porosity which, in turn, modulates the advective velocity via (\ref{eq:nvshperiodic}). It is usual to generate continuous Darcy flux distributions by interpolating between the finite-difference nodal fluxes. If, however, our interpolated estimate of the steady Darcy flux component is not precisely divergence-free, i.e. $\nabla \cdot \textbf{q}_{s} = \epsilon \neq 0$, where $\epsilon$ is independent of time, then the continuity equation becomes
\begin{equation} \label{:app_C3}
\dfrac{\partial \varphi}{\partial t} = - \nabla \cdot (\textbf{q}_{p} \, e^{i \omega t}) - \epsilon
\end{equation}
which provides for unbounded growth of $\varphi$ as $t \rightarrow \infty$, resulting in unphysical flow paths. To remedy this situation we impose equations (\ref{eqn:app_C2}) as exact constraints on the interpolated numerical solutions for $\textbf{q}$ and $\varphi$ via the following three-step process. 

\subsection*{Step 1: Enforcing zero divergence for $\textbf{q}_{s}$ via a streamfunction approach}
Consider the steady component of the Darcy flux, $\textbf{q}_s$, with nodal values $\textbf{q}_{s;m,n} = (\textbf{q}^x_{s; m,n},\textbf{q}^y_{s; m,n})$ and continuous vector interpolation $\textbf{q}^{\text{int}}_{s}(x,y) = (w_{x}(x,y),w_{y}(x,y))$. Even if $\textbf{q}^{\text{int}}_{s}$ exactly reproduces the numerical fluxes at the finite-difference node points it does not follow that the interpolation necessarily provides zero divergence at all locations $(x,y)$. We address this by constructing an effective streamfunction, $\Psi$, for $\textbf{q}_s$ based on $\textbf{q}^{\text{int}}_{s}$ and then differentiating $\Psi$ to generate modified flux components $\textbf{q}^{\ast}_{s, x}$ and $\textbf{q}^{\ast}_{s, y}$
according to
\begin{equation} \label{eqn:sfidentity_C4}
\textbf{q}^{\ast}_{s, x} \equiv \dfrac{\partial \Psi}{\partial y} \enskip , \quad \textbf{q}^{\ast}_{s, y} \equiv -\dfrac{\partial \Psi}{\partial x}.
\end{equation}
This results in an explicitly divergence-free flux distribution since 
\begin{equation} \label{eqn:divfree_C5}
\nabla \cdot (\textbf{q}^{\ast}_{s, x},\textbf{q}^{\ast}_{s, y}) \equiv \dfrac{\partial^2 \Psi}{\partial x \,  \partial y} - \dfrac{\partial^2 \Psi}{\partial y \, \partial x} = 0.
\end{equation}
The effective streamfunction is calculated by integrating the interpolation $\textbf{q}^{\text{int}}_{s}$ according to (\ref{eqn:sfidentity_C4}) over the finite difference grid, yielding the two datasets:
\begin{eqnarray} \label{eqn:effsf_C6}
\Psi^{(x)}_{m,n} &=& \int_{y_0}^{y_n}q^{\text{int}}_{s,x}(x_0+m\Delta,y^\prime)dy^\prime,\nonumber\\
\Psi^{(y)}_{m,n} &=& -\int_{x_0}^{x_m}q^{\text{int}}_{s,y}(x^\prime,y_0+n\Delta)dx^\prime.
\end{eqnarray}
where each integral is performed by quadrature over $N_{knots}$ points. These datasets are combined to provide two independent estimates of the streamfunction $\Psi$ over the finite difference grid as
\begin{eqnarray} \label{eqn:Psi1Psi2}
\Psi^{(1)}_{m,n} &=& \Psi^{(x)}_{m,n}+\Psi^{(y)}_{m,0},\nonumber\\
\Psi^{(2)}_{m,n} &=& \Psi^{(y)}_{m,n}+\Psi^{(x)}_{0,n}.
\end{eqnarray}
Rather than choose between these estimates, we assemble
the streamfunction in discrete form by the simple average $\Psi_{m,n} = (\Psi^{(1)}_{m,n} + \Psi^{(2)}_{m,n})/2$. The discrete streamfunction $\Psi_{m,n}$ is then converted to final continuous form ($\Psi(x,y)$) via interpolation and the modified flux components calculated via (\ref{eqn:sfidentity_C4}).

\subsection*{Step 2: Enforcing the continuity equation on $\textbf{q}_{p}$ and $\varphi_{p}$}

In order to ensure that the transient form of the continuity equation
(\ref{eqn:app_C2}) is also satisfied, we fix the modified transient
porosity component $\varphi_{p}^{\ast}$ by setting
$\varphi^{\ast}_{p}(x,y) = -(i \omega)^{-1} \nabla \cdot
\textbf{q}^{\text{int}}_{p}(x,y)$,
where $\textbf{q}^{\text{int}}_{p}$ is the interpolated transient flux
solution based on the finite-difference nodal values for
$\textbf{q}_{p}$.  The total modified Darcy flux is constructed as
$\textbf{q}^{\ast}(x,y,t) = \textbf{q}^{\ast}_{s}(x,y) +
\textbf{q}^{\text{int}}_{p}(x,y)\, e^{i \omega t}$
and the total modified porosity as
$\varphi^{\ast}(x,y,t) = \varphi^{\text{int}}_{s}(x,y) +
\varphi^{\ast}_{p}(x,y)\, e^{i \omega t}$.
In turn, the periodic head $h^{\ast}_{p}$ is fixed by
(\ref{eq:nvshperiodic},\ref{eq:periodicq}), yielding
$h^{\ast}(x,y,t) = h^{\text{int}}_{s}(x,y) + h^{\ast}_{p}(x,y)\, e^{i
  \omega t}$.
Thus, identically, we satisfy the continuity equation and
compressibility relation over the problem domain and obtain an exactly
mass-conserving velocity field
$\textbf{v}^{\ast} = \textbf{q}^{\ast}/\varphi^{\ast}$ that
incorporates the underlying tidal groundwater hydraulics.

\begin{figure}[t]
\centering
\includegraphics[scale=0.75]{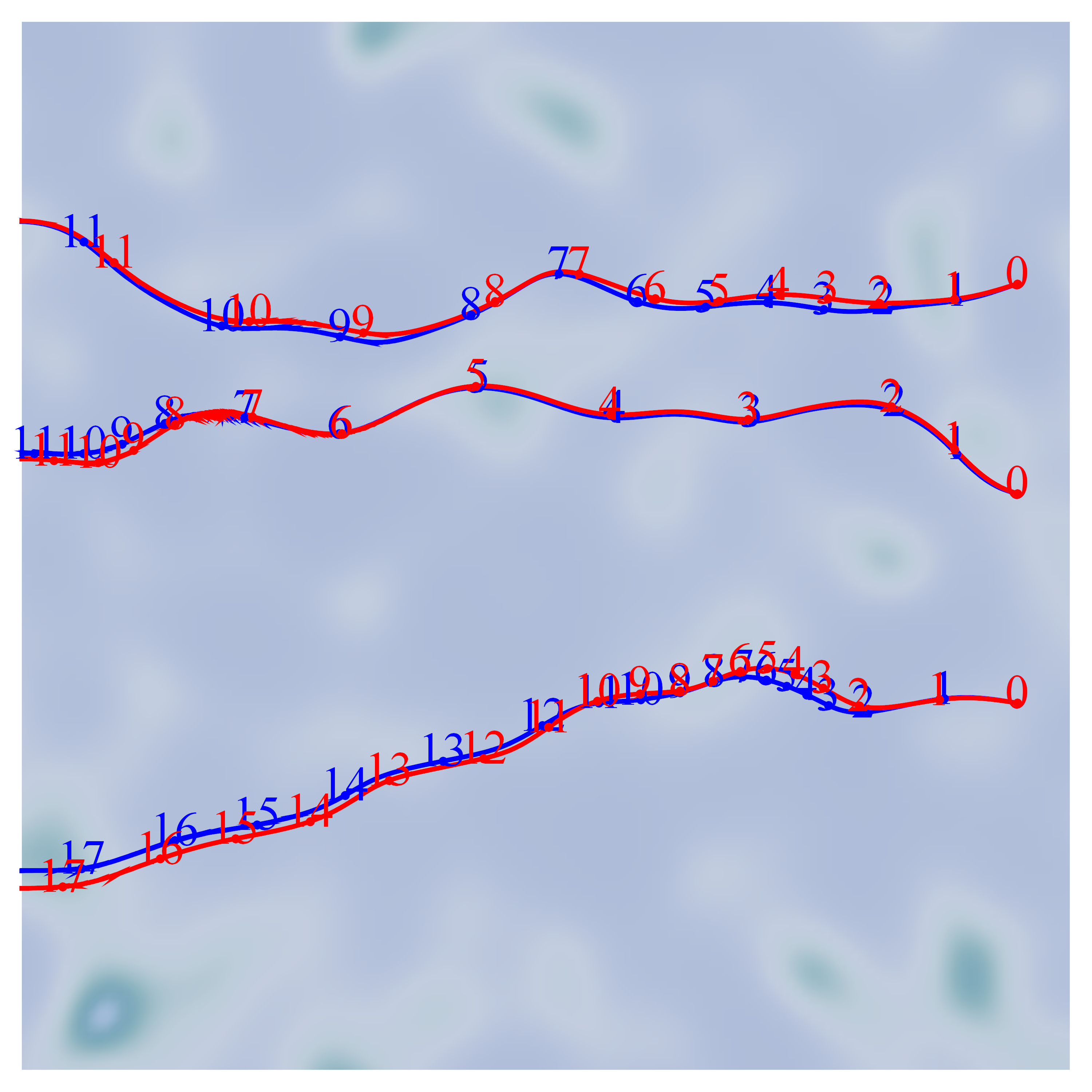}
\caption{Comparison of flow paths integrated forward in time using
  simple interpolation of finite-difference solutions (red) and full
  CE-corrected solutions (blue) superimposed on the associated
  $\kappa$ field (shaded).  Numbering refers to the number of elapsed
  periods \textit{P} ($\times 100$) along each path, starting at
  $t = 0$ at the points to the right.}
\label{fig:CEflowpathcomps}
\end{figure}

In general the updated variables differ from those interpolated
directly from the finite-difference flux solution but retain the
character of the underlying numerical solution.
Figure~\ref{fig:CEflowpathcomps} compares three arbitrary flow paths
calculated using $\textbf{v}^{\text{int}}$ with the counterpart path
calculated using $\textbf{v}^{\ast}$ with $N_{knots}=501$.

\subsection*{Step 3: Checking self-consistency of the adjusted dependent variables}
The previous two steps have ensured that the updated dependent variables ($h^{\ast},\textbf{q}^{\ast},\varphi^{\ast}$) satisfy the continuity equation and the compressibility relation, but at the cost of perturbing the variables from the finite-difference values. We assess the nature and magnitude of the perturbations using the Darcy equation (\ref{eq:Darcyq}) as a reference. The updated heads and fluxes will be consistent with a Darcian process if the updated flux vector $\textbf{q}^{\ast}$ and the gradient of the updated head $\nabla h^{\ast}$ are aligned. This alignment is tested by calculating azimuths $\theta_{\nabla h^{\ast}} = \textrm{ArcTan}(\nabla h^{\ast}_{y}/\nabla h^{\ast}_{x})$ and $\theta_{q^{\ast}} = \textrm{ArcTan}(q^{\ast}_{y}/q^{\ast}_{x})$ of the updated steady flux and steady head gradient vectors, with zero azimuth chosen to lie in the mean flow direction (negative $x$ direction). Azimuth plots were compiled by sampling the updated head and flux distributions at the finite difference node locations and plotting the sample values against each other, and comparing this with the discrete solution counterpart. 

\begin{sidewaysfigure}
\centering
\includegraphics[scale=2]{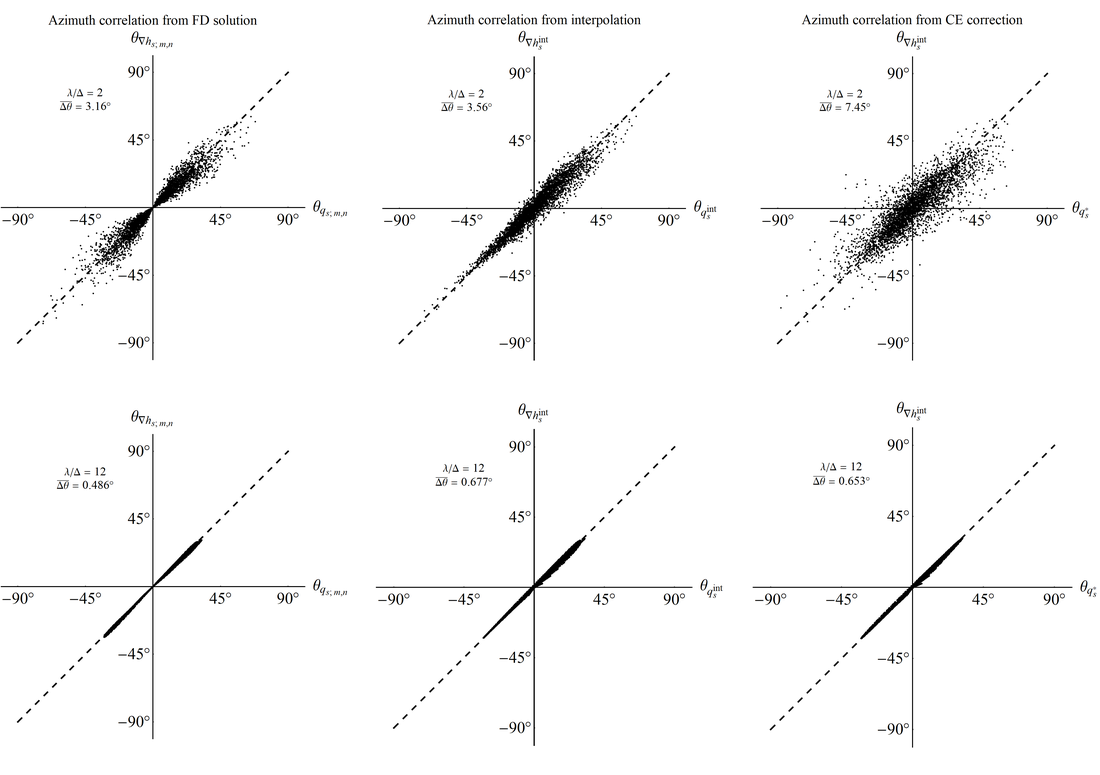}
\caption{Self-consistency checks of azimuths from steady fluxes and
  head gradients for the finite-difference solution (left column),
  interpolated solution (center column) and the updated flux (right
  column) evaluated for two different correlation lengths (upper and
  lower rows).  The numerical scheme gives better results at larger
  correlation lengths.}
\label{fig:CEazimuthcomps}
\end{sidewaysfigure}

\begin{figure}
\centering
\includegraphics[scale=1.2]{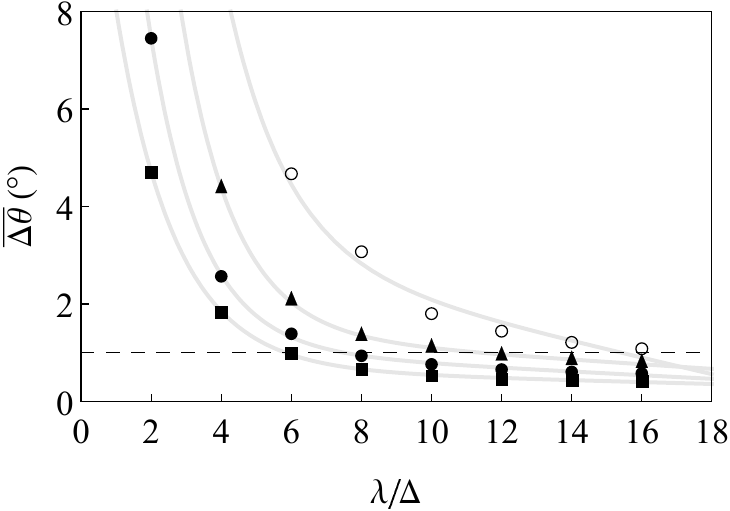}
\caption{Dependence of mean azimuthal difference $\overline{\Delta \theta}$ for the updated fluxes on the correlation length $\lambda/\Delta$ of a set of realizations of the input conductivity field with variances $\sigma^{2}_{\textrm{log}\kappa}$ = 1/2 (squares), 1 (dots), 2 (triangles) and 4 (circles). Curves are fitted to each set of points to aid the eye, and the acceptability cutoff is set at $1^{\circ}$ (dashed line).} 
\label{fig:CEazlambda}
\end{figure}

Results of this assessment are shown in Figure \ref{fig:CEazimuthcomps} for two different spatial correlation lengths $\lambda/\Delta$ and $(N_{x},N_{y}) = (64,64)$. In each case the finite-difference solution provides an azimuth plot with non-zero mean discrepancy between the head gradient and flux vectors, which is a consequence of the discrete heads being evaluated at grid nodes (with specified conductivity values) and the discrete fluxes at mid-nodal locations (with spatially averaged conductivity values). The finite-difference azimuth plots provide a lower bound to the error of the updated azimuth distributions. For $\lambda/\Delta = 2$ the updated azimuth plot shows wide dispersion, including many samples with opposed azimuths (approximately 15\% of the flux and head gradient vectors have opposite flow directions); however, for $\lambda/\Delta = 12$ the updated azimuths are more tightly aligned ($<1.4\%$ opposed flows). Figure \ref{fig:CEazlambda} displays the $\lambda$-dependence of the mean azimuthal discrepancy, showing that the mean azimuth error drops to less than $1^{\circ}$ for $\lambda/\Delta \geq 8$ for $\sigma^{2}_{\textrm{log}\kappa} = 1$. The results also depend on the conductivity variance $\sigma^{2}_{\textrm{log}\kappa}$: increasing variances worsen the azimuthal discrepancy.

Fundamentally, the results presented in this Appendix give confidence that our numerical scheme generates Darcian solutions that strictly obey the necessary continuity and compressibility constraints so long as the spatial conductivity variations are sufficiently well sampled (resolved) by the discrete set of input $\kappa$ values. In order to guarantee acceptable quantitative performance of the physical constraints and laws in our simulations it is useful to adopt a maximum permissible mean azimuthal discrepancy, e.g. $1^{\circ}$. In this case, for unit variance $\sigma^{2}_{\textrm{log}\kappa} = 1$ attention would be confined to cases where $\lambda/\Delta \geq 8$ as indicated in Figure \ref{fig:CEazlambda} (see dashed line).

\section{Connection of flow reversal to matrix conductivity and poroelasticity} \label{subsec:vorticity}
It is clear from Figure \ref{fig:psectionsx2} that the effect of the tidal boundary is to generate significant flow reversals that stimulate the Lagrangian phenomena of interest. Vorticity provides a measure of flow reversal and can be related to key physical processes in the poroelastic system.  Consider the vorticity $\bm{\nu}$ of our 2D groundwater flow, defined by
$\bm{\nu} \equiv \mathbf{\nabla} \times \mathbf{v} = \mathbf{\nabla} \times (\mathbf{q}/\varphi)$. By
a standard vector identity the curl of a product of a scalar and a vector is,
in the notation of our system,
\begin{equation}
\label{eq:vorticity1}
\boldsymbol{\nabla} \times \left(\frac{\mathbf{q}}{\varphi}\right) = 
 \nabla\left(\frac{1}{\varphi}\right) \times \mathbf{q} + 
 \left(\frac{1}{\varphi}\right) \boldsymbol{\nabla} \times \mathbf{q}
= 
 \varphi^{-2} \left(\nabla \varphi \times \mathbf{q}\right) + 
 \varphi^{-1} \left(\boldsymbol{\nabla} \times \mathbf{q}\right).
\end{equation}
Noting that $\nabla \varphi = S \nabla h$ (see equation (\ref{eq:nvsh})) and that the curl of the gradient of a scalar is zero results in the vorticity relation
\begin{equation}
\label{eq:vorticity2}
\bm{\nu} = \frac{\nabla h \times \nabla \kappa}{\varphi} 
\end{equation}
expressed in terms of the conductivity and porosity of the matrix.
Using the expression for the linear poroelasticity (\ref{eq:nvsh}) and defining $\hat{h}(\mathbf{x},t) \equiv (h(\mathbf{x},t)-h_{\textrm{ref}})/\textsl{g}_{p}$ as the non-dimensional local forcing, equation~(\ref{eq:vorticity2}) becomes
\begin{equation}
\label{eq:vorticity3}
\bm{\nu} = \frac{\bm{\nu}_{\kappa}}{1 + \mathcal{C} \hat{h}} 
\end{equation}
with
\begin{equation}
\bm{\nu}_{\kappa} = \frac{\nabla h \times \nabla \kappa}{\varphi_{\textrm{ref}}}
\end{equation}
the vorticity from shear flows generated solely by conductivity variations.

Equation~(\ref{eq:vorticity3}) implies that flow reversal occurs due
to conductivity variations and poroelasticity (if the storativity
varies in space, (\ref{eq:vorticity2}) includes an additional term
$\kappa h (\nabla h \times \nabla S)$ that could separately generate
vorticity due to non-trivial variation of the storage field). The
magnitude of $\bm{\nu}_{\kappa}$ is controlled by the conductivity
variance $\sigma_{\textrm{log} \kappa}^{2}$.  However, as discussed in
Appendix \ref{app:zero_compression}, it is the trajectory symmetry
breaking due to poroelasticity that produces interesting Lagrangian
transport structure. Flow reversal from non-trivial conductivity
variation and symmetry breaking from poroelasticity are the two crucial
physical ingredients in the creation of the transport dynamics we
observe.

\section{Fluid deformation gradient tensor and finite time Lyapunov exponent} \label{app:FTLE}

Chaotic advection in any fluid flow is characterised by the exponential stretching of fluid material elements, which gives rise to rapid growth of material interfaces, and hence accelerated mixing and transport. This phenomenon is typically characterised by the (infinite-time) Lyapunov exponent (\ref{eqn:inf_time_lyapunov}), but for particle orbits in open flows with finite residence time in the flow domain, it is more useful to characterize deformation in terms of the finite-time Lyapunov exponent (FTLE). To formally define and derive the FTLE $\Lambda$ and the associated fluid deformation gradient tensor $\mathbf{F}$, we first introduce the Lagrangian spatial coordinate system $\mathbf{X}$, which may be defined in terms of the Eulerian trajectory $\mathbf{x}(t,t_0;\mathbf{X})$ of a fluid particle at position $\mathbf{X}$ at time $t_0$. The Eulerian trajectory $\mathbf{x}(t,t_0;\mathbf{X})$ is a solution of the advection equation
\begin{equation} \label{eqn:Lagrangian}
\frac{\partial\mathbf{x}}{\partial t} = \mathbf v(\mathbf{x},t), \qquad \mathbf{x}(t=t_0,t;\mathbf{X})=\mathbf{X},
\end{equation}
and so  represents a transformation from Lagrangian ($\mathbf{X}$) to Eulerian ($\mathbf{x}$) spatial coordinates. The FTLE is computed from the fluid deformation gradient tensor $\mathbf{F}(t,t_0;\mathbf{X})$ which quantifies how the infinitesimal vector $d\mathbf{x}(t,t_0;\mathbf{X})$ deforms from its reference state $d\mathbf{x}(t=t_0,t_0;\mathbf{X}) = d\mathbf{X}$, and so represents the Jacobian tensor associated with a transformation from Lagrangian to Eulerian coordinates:
\begin{equation} \label{eqn:dxdX}
d\mathbf{x}=\mathbf{F}\cdot d\mathbf{X},
\end{equation}
and, equivalently,
\begin{equation} \label{eqn:Fdefn}
F_{ij}\equiv\frac{\partial x_i}{\partial X_j} \qquad \textrm{where} \qquad \mathbf{F} \equiv [F_{ij}]. 
\end{equation}
Following this definition, the deformation gradient tensor
$\mathbf{F}(t,t_0;\mathbf{X})$ evolves with travel time $t$ along a
Lagrangian trajectory (streamline) as
\begin{equation} \label{eqn:deform}
\frac{\partial\mathbf{F}}{\partial t}=\nabla\mathbf{v}\left[\mathbf{x}(t,t_0;\mathbf{X}),t\right]^\top\cdot\mathbf{F}(t,t_0;\mathbf{X}),\quad\mathbf{F}(t=t_0,t_0;\mathbf{X})=\mathbf{1},
\end{equation} 
where the superscript $\top$ denotes the transpose.  We refer to
(\ref{eqn:deform}) as the \emph{evolution equation} for $\mathbf{F}$.
Given computation of the deformation gradient tensor via, the FTLE is
then defined as
\begin{equation} \label{ftle}
\Lambda(t,t_0;\mathbf{X})\equiv\frac{1}{2(t-t_0)} \ln\nu_d,
\end{equation}
where $\nu_d$ is the largest eigenvalue of the right Cauchy-Green deformation tensor $\mathbf{C}$
\begin{equation} \label{cgtensor}
\mathbf{C} = \mathbf{F}(t,t_0;\mathbf{X})^\top\cdot \mathbf{F}(t,t_0;\mathbf{X}).
\end{equation}
Note that in the limit $t-t_0\rightarrow\infty$, the FTLE converges to the (infinite-time) Lyapunov exponent $\Lambda_\infty$, and due to ergodicity in chaotic regions, the ensemble average $\langle\Lambda(t,t_0;\mathbf{X})\rangle$ (where the average is performed over many starting positions $\mathbf{X}$) also converges to $\Lambda_\infty$. In this way, the infinite-time Lyapunov exponent in for chaotic regions in open flows may be accurately estimated even if the typical orbit residence time is short.

For incompressible flows, $\nabla\cdot\mathbf{v}=0$, (\ref{eqn:deform}) yields the result $\det(\mathbf{F})=1$, i.e. the volumes of fluid elements are preserved under the flow. However the groundwater velocity $\mathbf{v}=\mathbf{q}/\varphi$ is not divergence-free, so care needs to be taken in computing the fluid deformation gradient tensor $\mathbf{F}(t,t_0;\mathbf{X})$ and the associated FTLEs $\Lambda(t,t_0;\mathbf{X})$ in the linear groundwater flow (\ref{eq:gfe}) due to spatio-temporal variability of the porosity $\varphi(\mathbf{x},t)$. Conservation of fluid mass under the linear groundwater flow is captured explicitly by the continuity equation (\ref{eq:cedeform}) and tracking of fluid ``particles'' must be performed with respect to the groundwater velocity. Note that whilst the variable nature of $\varphi$ renders neither $\mathbf{v}$ nor $\mathbf{q}$ divergence-free, it is important that constraints imposed by (\ref{eq:cedeform}) on particle advection kinematics and fluid deformation are enforced so to eliminate spurious transport and mixing behaviours (see Appendix \ref{app:numerical_ce}). 
 
From the evolution equation (\ref{eqn:deform}), we find
\begin{equation} \label{eqn:detF_Lagrangian}
\det\mathbf{F}(t,t_0;\mathbf{X})=\exp\left[\int_{t_0}^t \nabla\cdot\mathbf{v}(\mathbf{x}(t^\prime,t_0;\mathbf{X}),t^\prime)\,dt^\prime \right],
\end{equation}
where the integral is also applied in fixed Lagrangian position $\mathbf{X}$. We derive an explicit expression for the divergence $\nabla\cdot\mathbf{v}$ as follows. We define the fluid material derivative as
\begin{equation} \label{eqn:fluid_mat_deriv}
\frac{D_f}{Dt}\equiv\frac{\partial}{\partial t}+\mathbf{v}\cdot\nabla,
\end{equation}
and so from (\ref{eq:cedeform}),
\begin{equation}
\nabla\cdot\mathbf{v}=-\frac{D_f}{Dt}\ln\varphi,
\end{equation}
hence
\begin{equation} \label{eqn:detF_varphi}
\det\mathbf{F}(t,t_0;\mathbf{X})=\exp\left[\int_{t_0}^t -\frac{D_f}{Dt^\prime}\ln\varphi\,\, dt^\prime\right]=\frac{\varphi(\mathbf{X},t_0)}{\varphi(\mathbf{x}(t,t_0;\mathbf{X}),t)}.
\end{equation}
The impact of mass conservation (\ref{eq:cedeform}) is that the volume of a moving fluid element scales inversely with the local porosity $\varphi$. Whilst this imposes temporal fluctuations in $\det\mathbf{F}$, these are transient and bounded, hence mass is conserved over arbitrarily long times. As such (\ref{eqn:detF_varphi}) serves as a consistency check during particle tracking and computation of FTLEs.

\section{Integrability of particle trajectories for incompressible media}
\label{app:zero_compression}

To show that fluid particle trajectories are \emph{integrable} for incompressible media (i.e. they cannot exhibit chaotic dynamics), we consider the linear groundwater equation (\ref{eq:gfe}) in the limit $S=0$ in terms of the steady $h_s(\mathbf{x})$ and periodic $h_p(\mathbf{x},t)$ head solutions
\begin{equation} \label{eqn:incomp_gfe}
\nabla\cdot\left(K(\mathbf{x})\nabla [h_s(\mathbf{x})+h_p(\mathbf{x},t)]\right)=0,
\end{equation}
subject to the fixed head boundary conditions $h|_{x=L}=\textsl{g}_L$, $h|_{x=0}=\textsl{g}(t)$, the latter of which may be decomposed into steady and zero-mean fluctuating components as $\textsl{g}(t)=\bar{\textsl{g}}+\textsl{g}^\prime(t)$. Due to linearity (\ref{eqn:incomp_gfe}) is satisfied individually by the steady and periodic solutions, which may be expressed as
\begin{eqnarray}
h_s(\mathbf{x}) &=& \textsl{g}_L+(\bar{\textsl{g}}-\textsl{g}_L)f(\mathbf{x}),\\
h_p(\mathbf{x},t) &=& \textsl{g}^\prime(t)f(\mathbf{x}),
\end{eqnarray}
where $f(\mathbf{x})$ is the solution to (\ref{eqn:incomp_gfe}) subject to the fixed head boundary conditions $h|_{x=L}=1$, $h|_{x=0}=0$. As the porosity $\varphi=\varphi_{\textrm{ref}}$ is constant in the limit $S=0$, from (\ref{eq:vsingle}) both the fluid velocity and Darcy flux are separable in space and time as
\begin{eqnarray} \label{eqn:sep_velocity}
\mathbf{v}(\mathbf{x},t) &=& \frac{1}{\varphi_{\textrm{ref}}}\mathbf{q}(\mathbf{x},t)=\frac{1}{\varphi_{\textrm{ref}}}K(\mathbf{x})\nabla h(\mathbf{x},t),\\
&=& \mathbf{v}_0(\mathbf{x})G(t),
\end{eqnarray}
where $\mathbf{v}_0(\mathbf{x})=K(\mathbf{x})\nabla f(\mathbf{x})$, $G(t)=\bar{\textsl{g}}-\textsl{g}_L+\textsl{g}^\prime(t)$. The evolution of position of fluid particles is then given by the steady 2D advection equation
\begin{equation} \label{eqn:2Dsteady_advect}
\frac{d\mathbf{x}}{d\tau}=\mathbf{v}_0(\mathbf{x}),
\end{equation}
where $d\tau=G(t)dt$.  Because (\ref{eqn:2Dsteady_advect}) is a steady
2D vector field, it does not admit chaos \cite{Aref_frontiers_2017}.
Note that as the porosity $\varphi_{\textrm{ref}}$ is constant, the
fluid velocity may be described in terms of the streamfunction
$\psi_0$ as $\mathbf{v}_0=\nabla\times\psi_0(\mathbf{x})\mathbf{e}_z$.
Hence for incompressible media ($S=0$), fluid particle trajectories
are confined to fixed one-dimensional ``streamlines'' which correspond to level
sets of the streamfunction $\psi_0$, even though the flow is unsteady.
Evolution along these ``streamlines'' is not monotone, but may
oscillate according to the forcing $G(t)$.  This is evident in Figure
\ref{fig:flowpaths} where flow paths calculated from a compressible
flow solution ($S > 0$) are compared with flow paths calculated for
incompressible flow ($S$ = 0).  The former paths display flow reversal
loops and braiding dynamics, while the latter resemble laminar
streamlines, inherently excluding chaos.  Compressibility of the
porous medium ($S>0$) is a necessary condition for the attainment of
chaotic dynamics.

\bibliographystyle{plainnat}
\bibliography{database.ref}

\end{document}